\documentclass[twocolumn,prb,aps,superscriptaddress,longbibliography]{revtex4-2}
\usepackage{amsmath}
\usepackage{amssymb}
\usepackage{amsfonts}
\usepackage[dvips]{graphicx}
\usepackage{subfigure}
\usepackage{dcolumn}
\usepackage{txfonts}
\usepackage{bm}
\usepackage{makeidx}
\usepackage{color}
\usepackage{mathtools}
\usepackage{threeparttable}
\usepackage{physics}
\usepackage[colorlinks,linkcolor=blue,anchorcolor=blue,citecolor=blue,urlcolor=blue]{hyperref}
\usepackage{lipsum}
\usepackage[table,xcdraw]{xcolor}
\usepackage[normalem]{ulem}

\begin{document}

\title{Sensitive dependence of Poor Man's Majorana modes on the length of the superconductor}

\author{Zhi-Lei Zhang}
\affiliation{Graduate School of China Academy of Engineering Physics, Beijing 100193, China}

\author{Xin Yue}
\affiliation{Beijing Computational Science Research Center, Beijing 100193, China}

\author{Guo-Jian Qiao} 
\affiliation{Graduate School of China Academy of Engineering Physics, Beijing 100193, China}

\author{C. P. Sun}
\email{suncp@gscaep.ac.cn}
\affiliation{Graduate School of China Academy of Engineering Physics, Beijing 100193, China}
 
\begin{abstract}
In a hybrid system where two quantum dots (QDs) are coupled to a conventional $s$-wave superconductor, Poor Man's Majorana modes (PMMs) have been proposed. Existing theories often idealize the superconductor (SC) as a bulk system or an infinitely long chain, or treat it as another QD with proximity-induced superconductivity, while experiments employ superconducting segments of finite length. Here, we model the SC as a finite-length 1D chain and treat the QDs and SC on equal footing. We obtain the conditions for the existence of PMMs, valid for arbitrary SC length and applicable to arbitrary tunneling strengths and magnetic fields. We find that the number of PMMs is highly sensitive to the SC length: it oscillates between zero and two with a period set by the Fermi wavelength ($\sim1\,\text{\AA}$), while four PMMs appear in the long-SC limit where the effective coupling between the two QDs becomes negligible. We further demonstrate that the PMMs that are separately localized at the two ends of the hybrid system do not exist in the finite-length case. Consequently, only nearly localized PMMs can be identified when the magnetic field is strong enough. In this way, the generalized `sweet spot' of the practical system can be found.
\end{abstract}

\maketitle

\section{Introduction\label{sec: appendix1}}
Majorana fermions in condensed matter systems, known as Majorana quasiparticles, have attracted significant interest owing to their non-Abelian statistics and potential applications in topological quantum computation~\cite{kitaev2001unpaired, PhysRevLett.104.040502, alicea2012new, lutchyn2018majorana, prada2020andreev}. In early studies, two classes of superconducting heterostructures—semiconductor nanowires (or topological insulators) in proximity to an \emph{s}-wave superconductor—emerged as the most promising platforms for realizing such quasiparticles~\cite{PhysRevLett.105.077001, PhysRevLett.105.177002, flensberg2021engineered, PhysRevLett.100.096407, PhysRevB.82.184516, Wang_2015, Chung_2011, 2025PRB, Qiao_2022, Yue2023, Qiao2024, yue2025finite, qiao2025size}. In these heterostructures, possible experimental signatures attributed to Majorana quasiparticles have been reported; however, no conclusive evidence has been found. This is because trivial mechanisms—such as disorder, smooth confinement, and inhomogeneous chemical potentials—can produce signatures indistinguishable from those of Majorana quasiparticles~\cite{prada2020andreev, PhysRevB.86.100503, PhysRevB.96.075161, Sankar_Das_Sarma2023}. To avoid these effects, minimal hybrid systems consisting of two quantum dots (QDs) coupled via a superconductor have attracted considerable attention due to their high tunability ~\cite{Flensberg2012, luethi2024perfect, liu2024enhancing, zhang2025poor, alvarado2026optimal}.

When subjected to strong magnetic fields, the electron spins become fully polarized, and the coupling between the QDs is mediated by the superconducting segment. As a result, the two-QD–SC hybrid system is considered to be effectively modelled by a spinless double quantum dot model, i.e., the `poor man's' Majorana model~\cite{Flensberg2012}. Based on this idealized model, two Majorana zero modes localized on the two QDs, as Majorana quasiparticle excitations, are predicted to appear only at specific parameter points (`sweet spots') and are therefore referred to as poor man’s Majorana modes (PMMs). More recently, motivated by numerous experimental signatures possibly associated with PMMs in realistic two-QD–SC systems, theoretical studies have focused on the conditions for their emergence. In these studies, the superconductor is typically modeled either as a single QD with proximity-induced superconductivity~\cite{liu2024enhancing, Creating2022} or as a bulk (or an infinite chain)~\cite{luethi2024perfect, zhang2025poor}. Remarkably, by eliminating quasiparticle excitations in the bulk superconductor, effective couplings between two QDs are induced via crossed Andreev reflection (CAR) and elastic cotunneling (ECT). The resulting low-energy theory shows that PMMs fully localized on the QDs are absent~\cite{luethi2024perfect, fateofDPPM}. Within a dressed approach that simultaneously treats quasiparticle excitations in SC and QDs on equal footing, our previous work~\cite{zhang2025poor} defined fully localized PMMs in the limit of an infinitely long SC and showed that four such PMMs can emerge, deviating from the idealized model, which predicts only two. These results indicate a discrepancy between the predictions based on the idealized model and current theoretical results.

In fact, the superconducting segment has a finite length ($\sim 300\,\mathrm{nm}$) in realistic experimental devices~\cite{dvir2023realization, ten2024two, zatelli2024robust, PhysRevX.13.031031, van2026single}, and thus the superconducting models currently employed are overly simplified. Intuitively, the length of SC directly controls the strength of SC-mediated processes, such as CAR and ECT. As the SC length increases, these induced couplings are progressively suppressed~\cite{PhysRevLett.111.060501, PhysRevLett.129.267701}, and they vanish in the limit of an infinitely long SC. Moreover, recent studies have shown that in other superconducting hybrid system, including topological insulator–SC~\cite{yue2025finite} and nanowire–SC~\cite{onedimensionSC, PhysRevLett.106.127001, qiao2025size}, the number of Majorana modes is extremely sensitive to the length of SC due to finite-length effects. Therefore, to resolve the aforementioned theoretical discrepancies and quantitatively determine the physical origin of the observed experimental signals, it is essential to investigate the finite-length effects of the SC and their influence on both the number and spatial distribution of PMMs. 

\begin{figure}
    \centering
    \includegraphics[width=8cm]{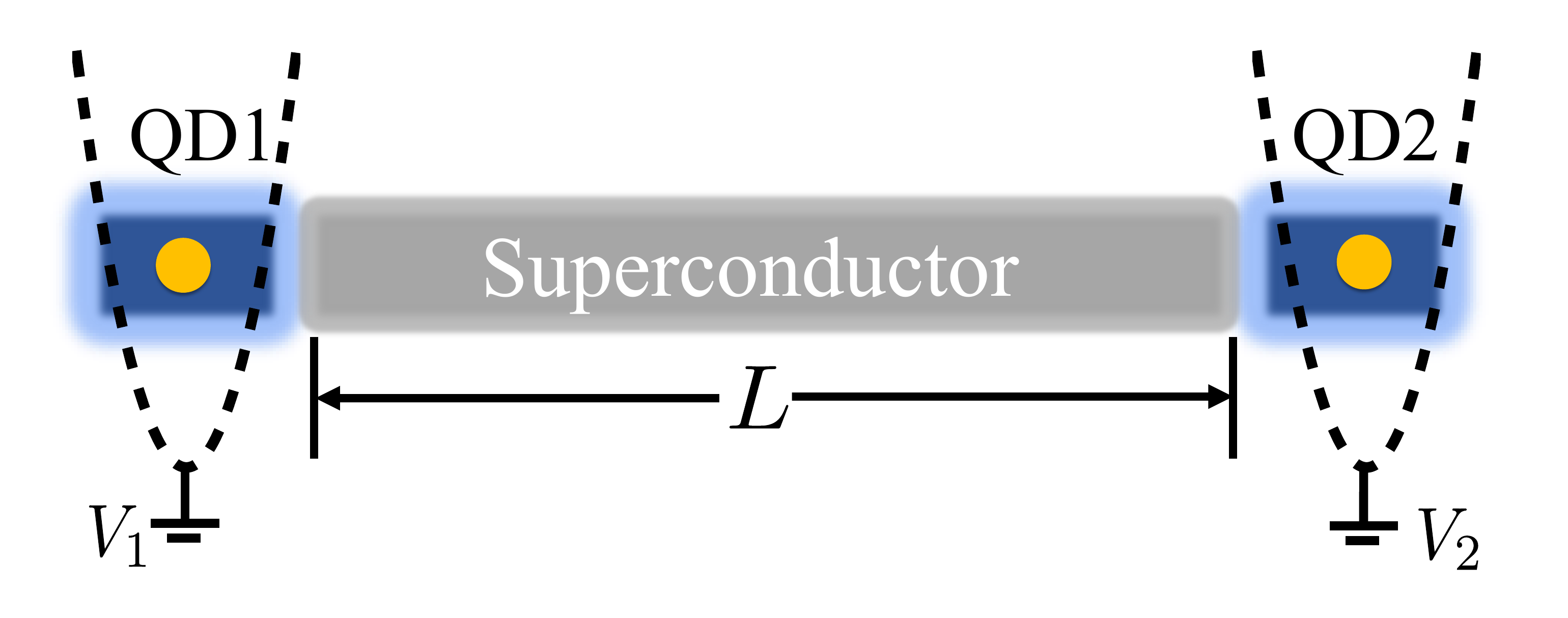}   \caption{Schematic of the hybrid system, where the length of the SC is denoted by $L$, and the chemical potentials of the two QDs can be tuned by the gate voltages $V_i$ ($i = 1, 2$) in the experiment.} \label{sketch map}
\end{figure}

In this paper, we investigate these issues by modeling the SC as a finite-length 1D chain. Starting from this microscopic model of QD-SC-QD (see Fig. \ref{sketch map}), we first derive the low-energy effective Hamiltonian for the two QDs and obtain the local and nonlocal couplings between the two QDs induced by SC, including the shift of chemical potential, the induced local pairing gap, the interdot hopping, the nonlocal pairing, and the effective spin-flip coupling. We show that both the interdot couplings~\cite{PhysRevLett.111.060501, PhysRevLett.129.267701} and the local couplings on each QD oscillate strongly with the SC length. Their oscillation period is on the order of the Fermi wavelength ($\sim 1\,\text{\AA}$), and their amplitudes decay exponentially as the SC length increases, with a decay scale set by the coherence length of the SC~\cite{yue2025finite, qiao2025size}. In the strong-magnetic-field limit, by integrating out the spin-up sector, we recover the ideal model~\cite{Flensberg2012} from the realistic microscopic description and obtain its corresponding existence condition.

By treating the quasiparticle excitations in both the QDs and the SC on an equal footing~\cite{Qiao2024, yue2025finite, zhang2025poor}, we then analytically derive the existence condition for PMMs in the hybrid system. This condition is valid for arbitrary SC length and over a broad range of magnetic fields and tunneling strengths, since the dressed effect of the SC is explicitly taken into account~\cite{Qiao2024,yue2025finite,zhang2025poor}. We show that the condition reduces to the known result in the long-SC limit~\cite{zhang2025poor}, while extending the analysis to realistic finite-length SC. For the practical system, which consists of an Al/Pt segment with a length of about $300\rm nm$, we identify the experimentally relevant physical parameter in which PMMs can exist.

Since the number of PMMs is determined by the effective couplings, and these couplings oscillate rapidly with the superconducting length, the number of PMMs also depends very sensitively on the SC length. Specifically, when the superconducting segment is much longer than the coherence length, the SC-induced interdot couplings vanish, and the system supports four PMMs~\cite{zhang2025poor}. In contrast, for finite SC lengths, the number of PMMs oscillates between zero and two as a function of length, with a period set by the Fermi wavelength. This behavior is consistent with the finite-size oscillations reported in other hybrid systems~\cite{yue2025finite, qiao2025size}. This resolves the discrepancy in the number of PMMs between our previous work~\cite{zhang2025poor} and the ideal model~\cite{Flensberg2012}. Since the number of PMMs directly determines the experimental signatures, the SC length critically affects whether Majorana-related signals can be observed.

Finally, by determining the spatial distribution of the wave functions of the two PMMs in the hybrid system, we show that, for a finite superconducting segment, PMMs strictly localized at the two opposite ends of the system do not exist. In fact, the corresponding localization condition defines a nonlinear system of equations for the SC length, QD chemical potential, and Zeeman energy, but this system admits no solution. By contrast, when the SC length approaches infinity, all PMMs supported by the system become localized at the two opposite ends, consistent with our previous result~\cite{zhang2025poor}. In the strong-magnetic-field limit, the localization condition admits an approximate solution, and the ideal model~\cite{Flensberg2012} is approximately recovered from the realistic microscopic system. This allows us to identify a generalized ``sweet spot'' for the practical QD--SC--QD hybrid system. Based on these results, we further identify the parameter regime in which two PMMs can be nearly localized at opposite ends for finite SC length.

\section{Finite-length model of Poor man's Majorana\label{sec: appendix2}}
In this section, we consider a finite-length model for the quantum dot-superconductor-quantum dot (QD-SC-QD) hybrid system, where the s-wave SC is described by a one-dimensional lattice chain consisting of $N$ sites. In the short-chain limit $N=1$, the SC model reduces to a quantum dot (QD) with proximity-induced superconductivity~\cite{liu2024enhancing, Creating2022}, while in the long-chain limit, it recovers the model studied in our previous paper~\cite{zhang2025poor}. It therefore gives a general description of realistic QD--SC--QD hybrid systems. This allows us to study the length dependence of the SC-induced couplings between the two QDs and to derive the existence conditions for PMMs for arbitrary SC length in subsequent sections.

The Hamiltonian of the hybrid system \cite{luethi2024perfect, liu2024enhancing, sau2012realizing, Qiao_2025} is 
\begin{equation}
H = H_d + H_s + H_t,
\label{eq1}
\end{equation} 
where $H_d$ and $H_s$ represent the Hamiltonians of the two QDs and the SC, respectively. Explicitly,
\begin{equation}
\begin{aligned}
    H_{d} &= \sum_{i=1,2} (\mu_d + h_d) d_{i \uparrow}^{\dagger} d_{i \uparrow} + (\mu_d - h_d) d_{i \downarrow}^{\dagger} d_{i \downarrow},\\
    H_s &= \sum_{n=1,\sigma}^{N} \tilde{\mu}_s c_{n,\sigma}^{\dagger}c_{n,\sigma} + ( \frac{t_{s}}{2}c_{n,\sigma}^{\dagger}c_{n+1,\sigma} - \Delta_{s}c_{n\uparrow}^{\dagger}c_{n\downarrow}^{\dagger} + {\rm{h.c.}}),
\end{aligned}
\end{equation}
where the s-wave superconductor is described by a one-dimensional lattice model of finite size, consisting of $N$ sites. The tunneling Hamiltonian between QDs and SC is~\cite{fateofDPPM, liu2024enhancing, Creating2022, luethi2024perfect,Qiao_2025, zhang2025poor}
\begin{equation}
\begin{aligned}
 H_{t} &= \sum_{\sigma} T d_{1 \sigma}^{\dagger} c_{1 \sigma} + \alpha [d_{1 \uparrow}^{\dagger} c_{1 \downarrow} - d_{1 \downarrow}^{\dagger} c_{1 \uparrow}] + {\rm h.c.} \\
    &+ \sum_{\sigma} T d_{2 \sigma}^{\dagger} c_{N \sigma} + \alpha [d_{2 \downarrow}^{\dagger} c_{N \uparrow} - d_{2 \uparrow}^{\dagger} c_{N \downarrow}] + {\rm h.c.}\\
\end{aligned}
\end{equation}
Here, $c_{n\sigma} (c_{n\sigma}^{\dagger})$ and $d_{i\sigma} (d_{i\sigma}^{\dagger})$ represent the annihilation (creation) operator of the SC and QD with spin $\sigma$ on site $n$ or $i$, respectively. Moreover, $\mu_d$ and $h_d$ are the on-site energy and the Zeeman energy of two QDs, $T$ is the strength of tunneling between the QDs and SC, and $\alpha$ is the spin-flipping coupling strength. The chemical potential of QDs $\mu_d$ can be adjusted experimentally by the applied gate voltage $V_i$, as shown in Fig. \ref{sketch map}. $\tilde{\mu}_s = \mu_s - t_s \cos(\pi/(N+1)) $ is the effective chemical potential of SC, $\Delta_s$ is the pairing strength, $t_s$ is the hopping strength of the adjacent lattice sites in the SC. 

Under open boundary conditions~\cite{qiao2025size}, through a transformation 
\begin{equation}
c_{n\sigma} = \sqrt{\frac{2}{N+1}}\sum_{l=1}^N \sin\left(\frac{\pi l  n}{N+1}\right) c_{l\sigma},
\end{equation}
the Hamiltonian of SC and tunneling Hamiltonian become
\begin{equation}
    \begin{aligned}
        H_s &= \sum_{l,\sigma} \bar{\mu}_{s}(l) c_{l\sigma}^\dagger c_{l\sigma} - \Delta_s (c_{l\uparrow}^\dagger c_{l\downarrow}^\dagger + \rm{h.c.}),\\
        H_t &= \sum_{l\sigma} T_l d_{1\sigma}^\dagger c_{l\sigma} + \alpha_l (d_{1\uparrow}^\dagger c_{l\downarrow} - d_{1\downarrow}^\dagger c_{l\uparrow})\\
        &+ \sum_{l\sigma} \tilde{T}_l d_{2\sigma}^\dagger c_{l\sigma} + \tilde{\alpha}_l (d_{2\downarrow}^\dagger c_{l\uparrow} - d_{2\uparrow}^\dagger c_{l\downarrow}),
    \end{aligned}\label{H_SC_l}
\end{equation}
where $\bar{\mu}_{s}(l) = t_s\{\cos [\pi l/(N+1)] - \cos [\pi /(N+1)]\} + \mu_s$ denotes the kinetic energy of superconducting electrons. The effective tunneling strength between QD-1 and SC is given by 
\begin{equation}
o_l = o \sqrt{\frac{2}{N+1}}\sin\left(\frac{\pi l }{N+1}\right),
\end{equation}
where $o = T$ and $o=\alpha$ denote the strengths of spin-conserving and spin-flip tunneling, respectively. For QD-2, the corresponding tunneling strengths are $\tilde{o}_l=(-1)^{l+1} o_l $.

In earlier studies, the effect of finite length of the SC was mainly entered only through the effective coupling between the two QDs under the weak tunneling regime  \cite{PhysRevLett.111.060501,PhysRevLett.129.267701}. By contrast, how these effective couplings vary with SC length for arbitrary tunneling strength and quantitatively affect the formation of PMMs remains unexplored. In our recent studies of finite-size effects in topological-insulator--SC~\cite{yue2025finite} and nanowire--SC~\cite{qiao2025size} hybrid systems, it was found that the SC-induced couplings oscillate rapidly with the size of the superconductor. As a result, the number of Majorana modes depends very sensitively on the SC size. This observation suggests that the SC length may likewise play an important role in determining whether PMMs emerge in the QD--SC--QD hybrid system. In what follows, we revisit the SC-induced couplings between the two QDs for arbitrary superconducting length and tunneling strengths. Through analyzing their dependence on the SC length, it is found that when the SC length is much larger than the coherence length, our results reduce to the effective coupling obtained in Ref.~\cite{PhysRevLett.111.060501}. On this basis, we further investigate the existence conditions for PMMs and the oscillatory evolution of their number with the superconducting length.

\section{Low-Energy Theory for Two Coupled QDs}

In this section, we analyze the dependence of the coupling strength between the two QDs induced by SC on the superconducting length through the low-energy effective Hamiltonian~\cite{PhysRevB.81.241310, PhysRevB.96.014510, PhysRevB.84.144522,yue2025finite}. We show that these coupling strengths exhibit an oscillatory decay with the SC length, with an oscillation period of about $k_F^{-1}\sim 1\,\text{\AA}$. When the SC length is much larger than the superconducting coherence length, only the induced local gap tends to a finite constant, while all other induced couplings decay exponentially, where the decay length is the superconducting coherence length. We also show how the idealized model~\cite{Flensberg2012} can be obtained from the microscopic model, thereby clarifying the connection between the two descriptions and the conditions under which the ideal model is valid.

\subsection{Induced Coupling Between Two QDs}
The SC-induced effective couplings in the two QDs can be captured by a low-energy effective model. Such an effective Hamiltonian for the coupled two-QD system is usually derived by eliminating the quasiparticle excitations in SC~\cite{PhysRevB.81.241310, PhysRevB.96.014510,yue2025finite}. In the basis of $\mathbf{C}^\dagger = [\mathbf{d}^\dagger,\;\mathbf{d},\;\mathbf{c}^\dagger,\;\mathbf{c}]^T$, where $\mathbf{d}^\dagger = [d_{1\uparrow}^\dagger, d_{1\downarrow}, d_{2 \uparrow}^\dagger, d_{2\downarrow}]^T$ and $\mathbf{c}=[c_{1\uparrow}^\dagger,c_{1\downarrow},\ldots,c_{N\uparrow}^\dagger,c_{N\downarrow}]^{T}$, the eigenvalue equation of the hybrid system is
\begin{equation}
    \begin{bmatrix}
        \mathcal{H}_d & \mathcal{T}\\
        \mathcal{T}^\dagger &\mathcal{H}_s
    \end{bmatrix} \begin{bmatrix}
        \Psi_d\\
        \Psi_s
    \end{bmatrix} = E \begin{bmatrix}
        \Psi_d\\
        \Psi_s
    \end{bmatrix}. \label{equation_of_Psi_s_Psi_d}
\end{equation}
Here, $\mathcal{H}_{d(s)} = \mathbf{H}_{d(s)}\otimes\sigma_z$ denotes the matrix for the Hamiltonian of QDs (SC) in the Nambu basis. $\mathcal{T} \equiv \mathbf{T} \otimes \sigma_z + i \boldsymbol{\alpha}\otimes\sigma_y$, where $\mathbf{T}$ and $\boldsymbol{\alpha}$ denote the tunneling and spin-flip coupling matrices. Specifically, the explicit forms of $\mathbf{H}_{d}$, $\mathbf{H}_{s}$, $\mathbf{T}$ and $\boldsymbol{\alpha}$ are given in Appendix~\ref{appendixA}. Expressing $\Psi_s$ in terms of $\Psi_d$ using Eq. \eqref{equation_of_Psi_s_Psi_d}, we obtain the equation for $\Psi_d$:
\begin{equation}
    \left[\mathcal{H}_d + \mathcal{T}(E-\mathcal{H}_s)^{-1}\mathcal{T}^\dagger\right] \Psi_d = E \Psi_d.
    \label{eq8}
\end{equation} 
In the low-energy limit $E/\Delta_s\ll1$, we expand the energy-dependent term $(E-\mathcal{H}_s)^{-1}$ and retain only the first-order term $E/\Delta_s$, then Eq. \eqref{eq8} reduces to $\mathcal{H}_{\rm{eff}}\Psi_d = E \Psi_d$ where
\begin{equation}
\mathcal{H}_{\rm{eff}}=(\mathbf{1} + \mathcal{T} \mathcal{H}_s^{-2} \mathcal{T}^\dagger)^{-1}(\mathcal{H}_d - \mathcal{T} \mathcal{H}_s^{-1} \mathcal{T}^\dagger).
\end{equation}
Therefore, the low-energy effective Hamiltonian of two coupled QDs can be defined as
\begin{equation}
\mathbf{H}_{\rm{eff}} = [\mathbf{d}^\dagger, \mathbf{d} ]\cdot\mathcal{H}_{\rm{eff}}\cdot[\mathbf{d}, \mathbf{d}^\dagger ]^T,
\label{eq10}
\end{equation}
where $\mathcal{H}_{\mathrm{eff}}$ is the matrix representation of the effective Hamiltonian:
\begin{equation}
\mathcal{H}_{\rm{eff}}= \frac{1}{1+\zeta} \begin{bmatrix}
\mathbf{h}_d & \mathbf{p}_d\\
\mathbf{p}_d^\dagger & -\mathbf{h}_d
\end{bmatrix}.\label{eq_of_effective_QD}
\end{equation}
Here, $\mathbf{h}_d \equiv (\bar{\mu}_d \sigma_z + h_d \sigma_0 - \bar{\Delta}_s \sigma_x) \otimes \sigma_0 - (t \sigma_z + \Delta_{sp} \sigma_x)\otimes\sigma_x$ and $\mathbf{p}_d \equiv i (\Delta \sigma_z - \bar{\alpha}\sigma_x)\otimes\sigma_y$, where $\bar{\mu}_d \equiv \mu_d - \bar{\epsilon}$ and $1/(1+\zeta)$ is the renormalization factor with $\zeta \equiv \bar{\Delta}_s/\Delta_s$. 

In this effective description, by eliminating the virtual processes of the exchange of quasiparticle excitations in SC, both local and nonlocal couplings in the QD system are induced. The local terms are characterized by:
\begin{equation}
    \bar{\epsilon} = \sum_{i=1}^N \frac{\abs{T_i}^2 + \abs{\alpha_i}^2}{\epsilon_i^2 + \Delta_s^2} \epsilon_i,\;\bar{\Delta}_s = \sum_{i=1}^N \frac{\abs{T_i}^2 + \abs{\alpha_i}^2}{\epsilon_i^2 + \Delta_s^2} \Delta_s,\label{eq_epsilon_deltabar}
\end{equation}
where $\bar{\epsilon}$ describes the shift of the QD chemical potential caused by virtual tunneling processes through the SC, while $\bar{\Delta}_s$ represents the induced local pairing gap generated by local Andreev reflection, in which an electron in a QD is coherently converted into a hole on the same QD through the SC. The nonlocal terms arise from processes involving both two QDs. The effective hopping amplitude and effective spin-flipping coupling between the two QDs 
\begin{equation}
t = \sum_{i=1}^N \frac{T_i \tilde{T}_i - \alpha_i \tilde{\alpha}_i}{\epsilon_i^2 + \Delta_s^2} \epsilon_i,\quad
\bar{\alpha} = \sum_{i=1}^N \frac{\alpha_i \tilde{T}_i + T_i \tilde{\alpha}_i}{\epsilon_i^2 + \Delta_s^2} \epsilon_i,\label{eq_t_delta}
\end{equation}
are induced, respectively, by ECT and spin-dependent tunneling processes mediated by the superconducting segment. The pairing couplings $\Delta_{sp}$ and $\Delta$ are induced by CAR, in which electrons from the two different QDs combine into a Cooper pair in the SC, thereby generating an effective nonlocal pairing between the dots:
\begin{equation}
\begin{aligned}
\Delta_{sp} = \sum_{i=1}^N \frac{T_i \tilde{T}_i - \alpha_i \tilde{\alpha}_i}{\epsilon_i^2 + \Delta_s^2} \Delta_s,\;\;\Delta = \sum_{i=1}^N \frac{\alpha_i \tilde{T}_i + T_i \tilde{\alpha}_i}{\epsilon_i^2 + \Delta_s^2} \Delta_s.\label{eq_delta_sp_delta}
\end{aligned}
\end{equation}
Therefore, the finite superconductor not only renormalizes the local properties of each QD but also mediates coherent interdot tunneling processes, spin-flip processes, and electron pairing processes between them.

When the length of SC is greater than the coherence length ($L \gg \xi_0$), the chemical-potential shift on each QD and the local pairing gap are reduced to
\begin{equation}
    \bar{\epsilon} \approx 2 \Delta_0 \sin(2 k_F L) e^{- \frac{L}{\xi_0}},\;\bar{\Delta}_s \approx \Delta_0[1 + 2 \cos(2 k_F L) e^{-\frac{L}{\xi_0}}],
    \label{eq15}
\end{equation}
where $\Delta_0 = (T^2 + \alpha^2)a/(\Delta_s \xi_0)$. Here, $a$ is the lattice length, $k_F = \sqrt{2m_s\bar{\mu}_s}/\hbar$ is the effective Fermi wave vector and $\xi_0 = \hbar v_F/(2\Delta_s)$ is the coherence length of SC, where $\bar{\mu}_s \equiv \mu_s + \hbar^2 \pi^2/(2m_sL)$. We find that both couplings oscillate strongly with the superconducting length, with a period set by $k_F^{-1}\sim 1\,\text{\AA}$. At the same time, their amplitudes decay exponentially as the SC length increases. Notably, the shift of chemical potential decays to zero, whereas the induced local gap approaches a finite value $\Delta_0$. The exact expressions of these two couplings for arbitrary SC length are given in Appendix~\ref{appendixA}. Similarly, in the long-SC limit $L\gg \xi_0$, the SC-induced nonlocal couplings between the two QDs are obtained analytically as
\begin{equation}
\begin{aligned}
\Delta &\approx \Delta_1 [1 + 2 \cos(2 k_F L) e^{-\frac{L}{\xi_0}}]\cos(k_F L)e^{-\frac{L}{2\xi_0}},\\
\Delta_{sp} &\approx \Delta_2 [1 + 2 \cos(2 k_F L) e^{-\frac{L}{\xi_0}}]\cos(k_F L)e^{-\frac{L}{2\xi_0}},
\label{eq16}
\end{aligned}
\end{equation}
where $\Delta_1 = -4T\alpha a/(\Delta_s\xi_0)$ and $\Delta_2 = -2(T^2 - \alpha^2) a/(\Delta_s\xi_0)$ characterize the amplitudes of the two effective pairing. From these expressions, one can see that the nonlocal couplings oscillate with a period of about $1\,\text{\AA}$ and decay on the scale of $2\xi_0$. In the weak-tunneling regime, this length-dependent oscillatory behavior recovers the effective-coupling results obtained in Ref.~\cite{PhysRevLett.111.060501}. The effective spin-flip couplings and hopping between QDs are given by
\begin{equation}
\bar{\alpha} \approx \Delta_1 [1 + 2 \cos(2 k_F L) e^{-\frac{L}{\xi_0}}]\sin(k_F L)e^{-L/(2\xi_0)},\; t \approx \frac{\Delta_2}{\Delta_1} \bar{\alpha}.
\label{eq17}
\end{equation}
They exhibit the same oscillatory behavior with the superconducting length as the nonlocal pairing couplings. The exact expressions of these four nonlocal couplings for arbitrary SC length are also given in Appendix~\ref{appendixA}. In the limit $L\to\infty$, all these nonlocal couplings vanish. 

\begin{figure}
    \centering
    \includegraphics[width=8.5cm]{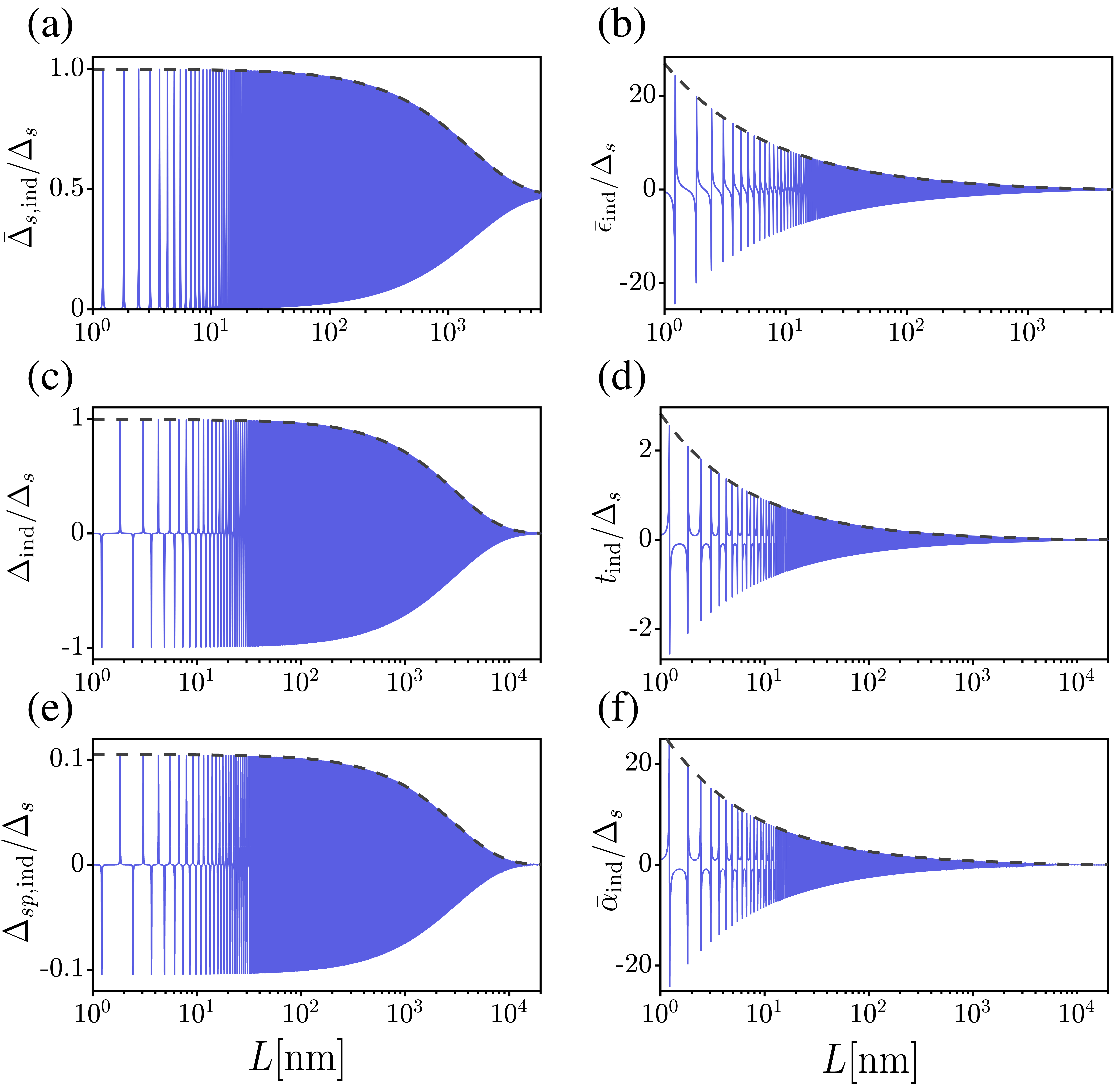}   \caption{The oscillatory dependence of SC-induced couplings between the two QDs on the superconducting length. All couplings oscillate with a period of about $1\,\text{\AA}$. The dashed lines denote the upper envelopes of the curves. When the SC length is much larger than the coherence length, only the induced gap remains finite, while all other induced couplings decay to zero. Panels (a) and (b) show the SC-induced local couplings on each QD: (a) the induced gap generated by Andreev reflection, and (b) the induced chemical-potential shift. Panels (c)–(f) show the SC-induced interdot couplings: (c) and (e) the pairing couplings induced by CAR, (d) the hopping coupling induced by ECT, and (f) the induced effective spin-flipping coupling. The parameters are $\Delta_s \approx 0.34{\rm{meV}}, \mu_s = 1\mathrm{eV}$, $t_s = 10 \mu_s$, $\xi_0\approx1600{\rm{nm}}$, $T \approx 32.6{\rm{meV}}$, $\alpha = 0.9 T$.}
    \label{oscillation}
\end{figure}

Since the experimentally observable quantities are ultimately determined by the renormalized effective couplings, we now turn to a detailed analysis of these induced couplings after renormalization. The renormalized induced couplings are defined by $o_{\rm ind}=o/(1+\zeta)$, where $o=t,\bar{\alpha},\Delta_{sp},\Delta,\bar{\epsilon},\bar{\Delta}_s$ and $\zeta\equiv \bar{\Delta}_s/\Delta_s$ ~\cite{yue2025finite, qiao2025size, PhysRevB.81.241310, PhysRevB.96.014510, PhysRevB.84.144522}. Their analytic forms have been derived in Eqs.~(\ref{shift_mu}-\ref{Delta_alpha}) in Appendix~\ref{appendixA}, and the corresponding renormalized one are numerically illustrated in Fig.~\ref{oscillation}. It is shown that all of these induced couplings exhibit rapid oscillations with a period of about $1\,\text{\AA}$, and the dashed curves mark the corresponding upper envelopes. Specifically, Fig. \ref{oscillation} (a) and (b) describe the SC-induced local couplings on each QD, including the induced gap generated by Andreev reflection and the induced shift of chemical potential. By contrast, panels (c)–(f) show the SC-induced interdot couplings, namely the pairing couplings induced through CAR, the hopping coupling induced through ECT, and the effective spin-flipping coupling generated by the SC. These results demonstrate that the finite length of the SC strongly affects both the local coupling on each QD and the effective coupling between the two QDs.

\subsection{Poor Man’s Model for a hybrid two QDs–SC system}
To reveal the connection between the realistic model and the ideal one~\cite{Flensberg2012}, we further derive the low-energy effective Hamiltonian for the spin-down sector. Under the spin-up basis $\mathbf{d}_\uparrow^\dagger \equiv [d_{1\uparrow}^\dagger,d_{2\uparrow}^\dagger,d_{1\uparrow},d_{2\uparrow}]$ and the spin-down basis $\mathbf{d}_\downarrow^\dagger \equiv [d_{1\downarrow}^\dagger,d_{2\downarrow}^\dagger,d_{1\downarrow},d_{2\downarrow}]$, the effective Hamiltonian in Eq. \eqref{eq10} is rewritten as 
\begin{equation}
    \mathbf{H}_{\rm{eff}} = \begin{bmatrix}
        \mathbf{d}_\uparrow^\dagger& \mathbf{d}_\downarrow^\dagger
    \end{bmatrix} \begin{bmatrix}
        \mathcal{H}_\uparrow & \mathcal{T}_{\uparrow \downarrow}\\
        \mathcal{T}_{\uparrow \downarrow}^\dagger & \mathcal{H}_\downarrow
    \end{bmatrix} \begin{bmatrix}
        \mathbf{d}_\uparrow\\
        \mathbf{d}_\downarrow
    \end{bmatrix},
\end{equation}
where $\mathcal{H}_\uparrow$ and $\mathcal{H}_\downarrow$ are the matrix representations of the spin-up and spin-down sectors, respectively, and $\mathcal{T}_{\uparrow\downarrow}$ describes the coupling between the two spin sectors. These matrices can be obtained straightforwardly from the matrix of the effective Hamiltonian of the two coupled QDs [Eq. \eqref{eq_of_effective_QD}]. When the magnetic field is applied along the positive $z$ direction ($h_d>0$), such that the low-energy excitations of the effective Hamiltonian satisfy
\begin{equation}
E \ll \min\!\left[E_\pm\right], \label{condition_effective_down}
\end{equation}
the spin-up sector can be regarded as a high-energy sector and can therefore be integrated out. Here, $E_\pm \equiv \sqrt{(\bar{\mu}_d+h_d)^2+\Delta^2} \pm t$ is the eigenvalue of the spin-up sector and they should be nonzero to keep Eq. \eqref{condition_effective_down} set up. Following the same procedure used to derive the low-energy effective Hamiltonian for the QDs, we obtain the effective Hamiltonian in the spin-down sector.
\begin{equation}
    H_{\downarrow,\rm{eff}} = \sum_{i=1,2} (\tilde{\mu}_d - \tilde{h}_d) d_{i\downarrow}^\dagger d_{i\downarrow} - \tilde{t} d_{1\downarrow}^\dagger d_{2\downarrow} - \tilde{\Delta}d_{1\downarrow}^\dagger d_{2\downarrow}^\dagger + \rm{h.c.} \label{practical_flensberg}
\end{equation}
Here, the spin-down sector acquires an effective chemical potential and the Zeeman field, 
\begin{equation}
    \begin{aligned}
        \tilde{\mu}_d(L) \equiv \chi_- \bar{\mu}_d + \frac{2\bar{\Delta}_s(\bar{\alpha} \Delta + t \Delta_{sp})}{Z},\quad \tilde{h}_d(L) \equiv \chi_+ h_d,
        \label{chiandxi}
\end{aligned}
\end{equation}
which describes the local on-site energy of spin-down electrons, where $\chi_{\pm} \equiv \left[1 \pm (\bar{\alpha}^2 - \bar{\Delta}_s^2 - \Delta_{sp}^2)/Z\right]$ and the correction factor is $Z \equiv E_+ E_-$. It is noticed that the correction factor is nonzero, i.e., $Z \neq 0$. Moreover, an effective hopping amplitude $\tilde{t}$ and an equal-spin pairing amplitude $\tilde{\Delta}$ between the two QDs are obtained as
\begin{equation}
    \begin{aligned}
        \tilde{t}(L) &\equiv \sigma_+ t - \frac{2\Delta_{sp}[\bar{\Delta}_s (\bar{\mu}_d + h_d) + \Delta \bar{\alpha}]}{Z},\\
        \tilde{\Delta}(L) &\equiv \sigma_- \Delta + \frac{2\bar{\alpha}[\bar{\Delta}_s (\bar{\mu}_d + h_d) + t \Delta_{sp}]}{Z}.
        \label{tanddelta}
    \end{aligned}
\end{equation}
where $\sigma_{\pm} = 1 \mp {(\bar{\alpha}^2 \pm \bar{\Delta}_s^2 + \Delta_{sp}^2)}/{Z}$. Physically, these effective parameters arise from virtual processes that exchange electrons with the spin-up sector. Because these couplings are determined by the SC-induced couplings, they inherit a similar oscillatory dependence on the superconducting length. Eq.~\eqref{practical_flensberg} is the low-energy effective Hamiltonian for the spin-down sector of the full microscopic hybrid system, in which the effects of the SC and the spin-up sector are encoded in the effective parameters. Here, we show that, only under strong spin polarization, subject to a magnetic field satisfying Eq.~\eqref{condition_effective_down}, can the realistic microscopic model be reduced to the idealized one~\cite{Flensberg2012}.

Above, we have demonstrated the oscillatory behavior of the SC-induced couplings between the two QDs. In the following, we analyze how these oscillations influence the existence conditions and the number of PMMs.

\section{Existence Conditions and Number of PMMs}

In this section, we derive the existence condition for PMMs in the QD--SC--QD hybrid system.  Because we treat the quasi-excitations in the QDs and the SC on an equal footing, the resulting condition remains valid for arbitrary SC lengths and applies over a broad range of magnetic fields and tunneling strengths. We show that it reduces to our previous result in the infinitely long-SC limit, while for finite SC lengths the PMM existence regime depends sensitively on the superconducting length through the oscillatory SC-induced couplings. Since the experimental signatures of PMMs, such as the zero-bias differential-conductance signal, are directly determined by their number, we then examine how the number of PMMs evolves with the SC length. In the long-SC limit, the system hosts four PMMs, whereas for finite SC lengths the number of PMMs oscillates rapidly between zero and two with a period of about $1\,\text{\AA}$. A similar oscillation period has also been reported in other hybrid systems with superconducting finite-length effects.

We diagonalize the Hamiltonian of the hybrid system [see Eq. \eqref{eq1}] in terms of Bogoliubov quasiparticles, $H = \sum_\nu E_{\nu} \gamma_{\nu}^{\dagger} \gamma_{\nu}$, where the annihilation operator of the quasiparticle is a linear combination of electron and hole operators in both the QDs and the SC:
\begin{equation}
    \gamma_{\nu} = \mathbf{u}_\nu^d \cdot \mathbf{d} + \mathbf{d}^{\dagger} \cdot (\mathbf{v}_\nu^d)^T + \mathbf{u}_\nu^s \cdot \mathbf{c} + \mathbf{c}^{\dagger} \cdot (\mathbf{v}_\nu^s)^T. \label{quasi-particle operation}
\end{equation}
Here, $\mathbf{d}^\dagger = [d_{1\uparrow}^\dagger, d_{1\downarrow}, d_{2\uparrow}^\dagger,  d_{2\downarrow}]^T$ denotes the electron operator of the two QDs, and $\mathbf{c}^\dagger = [c_{1\uparrow}^\dagger, c_{1\downarrow}, \ldots, c_{N\uparrow}^\dagger, c_{N\downarrow}]^T$ denotes those of the SC. The definitions of electron and hole wave functions are
\begin{align}
\mathbf{u}^{\nu}_\alpha &= [u_{\nu,1\uparrow}^\alpha, u_{\nu,1\downarrow}^\alpha, \dots, u_{\nu, N_\alpha\uparrow}^\alpha, u_{\nu, N_\alpha\downarrow}^\alpha],\\
\mathbf{v}^{\nu}_\alpha &= [v_{\nu,1\uparrow}^\alpha, v_{\nu,1\downarrow}^\alpha, \dots, v_{\nu, N_\alpha\uparrow}^\alpha, v_{\nu, N_\alpha\downarrow}^\alpha].\label{definition_of_u}
\end{align}
where $N_s = N$ represents the total number of sites in the SC ($\alpha = s$), and $N_d = 2$ corresponds to the QDs ($\alpha = d$).

PMMs are defined by the self-conjugation condition $\gamma_\nu = \gamma_\nu^{\dagger}$. This constraint implies that the electron and hole wave functions satisfy $\mathbf{u}^{\nu}_\alpha = (\mathbf{v}^{\nu}_\alpha)^*$ with $\alpha = d,s$. Owing to the particle–hole symmetry of the system~\cite{li2014probing, Qiao_2022, Qiao2024}, such a mode necessarily appears at zero energy, i.e., $E_\nu = 0$. Consequently, the existence of PMMs is determined by whether the following zero-energy equation admits a solution,
\begin{equation}
    \begin{aligned}
        \begin{bmatrix}\mathbf{h} & \mathbf{p}\\
\mathbf{p^{\dagger}} & -\mathbf{h}
\end{bmatrix} \begin{bmatrix}
    \mathbf{u}\\
    \mathbf{u}^*
\end{bmatrix} = 0,
\end{aligned}\label{zeroeigenfunction}
\end{equation} 
where $\mathbf{u} \equiv [\mathbf{u}_d\; \mathbf{u}_s]$, whose components correspond to the wave functions of the QD and the SC, respectively, and the subscript $\nu$ has been omitted. $\mathbf{h}$ and $\mathbf{p}$ are defined in Appendix \ref{appendixA}. By decomposing the wave functions of QD and SC into their real and imaginary parts: $\mathbf{u}_{d(s)} = \mathbf{u}^r_{d(s)} + i \mathbf{u}^i_{d(s)}$, we obtain the equations that $\mathbf{u}_d^r$ satisfies
\begin{equation}
\begin{aligned}
 \left[\mathbf{H}_d - (\mathbf{T} + \boldsymbol{\alpha})\cdot \mathbf{H}_s^{-1} \cdot(\mathbf{T} - \boldsymbol{\alpha})^\dagger\right] \cdot \mathbf{u}_d^r = 0.
\end{aligned}\label{eqofQD}
\end{equation}
Here, $\mathbf{T}$ and $\boldsymbol{\alpha}$ denote the tunneling and spin-flip coupling matrix, $\mathbf{H}_d$ and $\mathbf{H}_s$ represent the matrix representation of QD and SC, respectively. These matrices have been used in Eq. \eqref{equation_of_Psi_s_Psi_d}. To ensure the emergence of PMMs in the hybrid system, the determinant of the coefficient matrix in Eq.~\eqref{eqofQD} must be zero. This determinant equation determines the general condition for the existence of PMMs. When the correction factor is nonzero, $Z\neq 0$, this general criterion is reduced to the explicit existence condition for PMMs
\begin{equation}
\left[\tilde{\mu}_d(L) - \tilde{h}_d(L) \right]^2 - \tilde{t}(L)^2 + \tilde{\Delta}(L)^2 = 0, \label{sweet spot}
\end{equation}
where $\tilde{\mu}_d$, $\tilde{h}_d$, $\tilde{t}$ and $\tilde{\Delta}$ are defined in Eq. \eqref{chiandxi} and \eqref{tanddelta}. These quantities are precisely the effective couplings appearing in the low-energy Hamiltonian of the spin-down sector. Since they are functions of the SC-induced couplings, the existence condition for PMMs also becomes a function of the superconducting length. 

In the limit of an infinitely long SC, it follows from Eqs. (\ref{eq16}, \ref{eq17}) that the SC-induced pairing strength, hopping strength, effective spin-flipping coupling, and shift of chemical potential all vanish, and Eq.~\eqref{sweet spot} reduces to
\begin{equation}
    \begin{aligned}
        \mu_d^2 + \bar{\Delta}_s^2 - h_d^2 = 0
    \end{aligned}\label{Ltoinfty}
\end{equation}
where $\bar{\Delta}_s$ is the effective gap for each QD induced by SC. This condition is already shown in Ref. \cite{zhang2025poor}. More generally, away from this limiting case, the existence of PMMs in Eq.~\eqref{sweet spot} is dependent on the superconducting length. This is because these SC-induced parameters depend explicitly on the length of the SC. Due to the equal-footing treatment of quasiparticle excitations in both the QDs and the SC, this condition is valid for arbitrary magnetic fields and tunneling strength as long as these parameters do not disrupt the structures of SC or QD. It therefore directly identifies the physical parameter regime (i.e., chemical potential, magnetic field of QDs, and so on) in which PMMs can emerge for an SC with a given length.

\begin{figure}
    \centering
    \includegraphics[width=8.5cm]{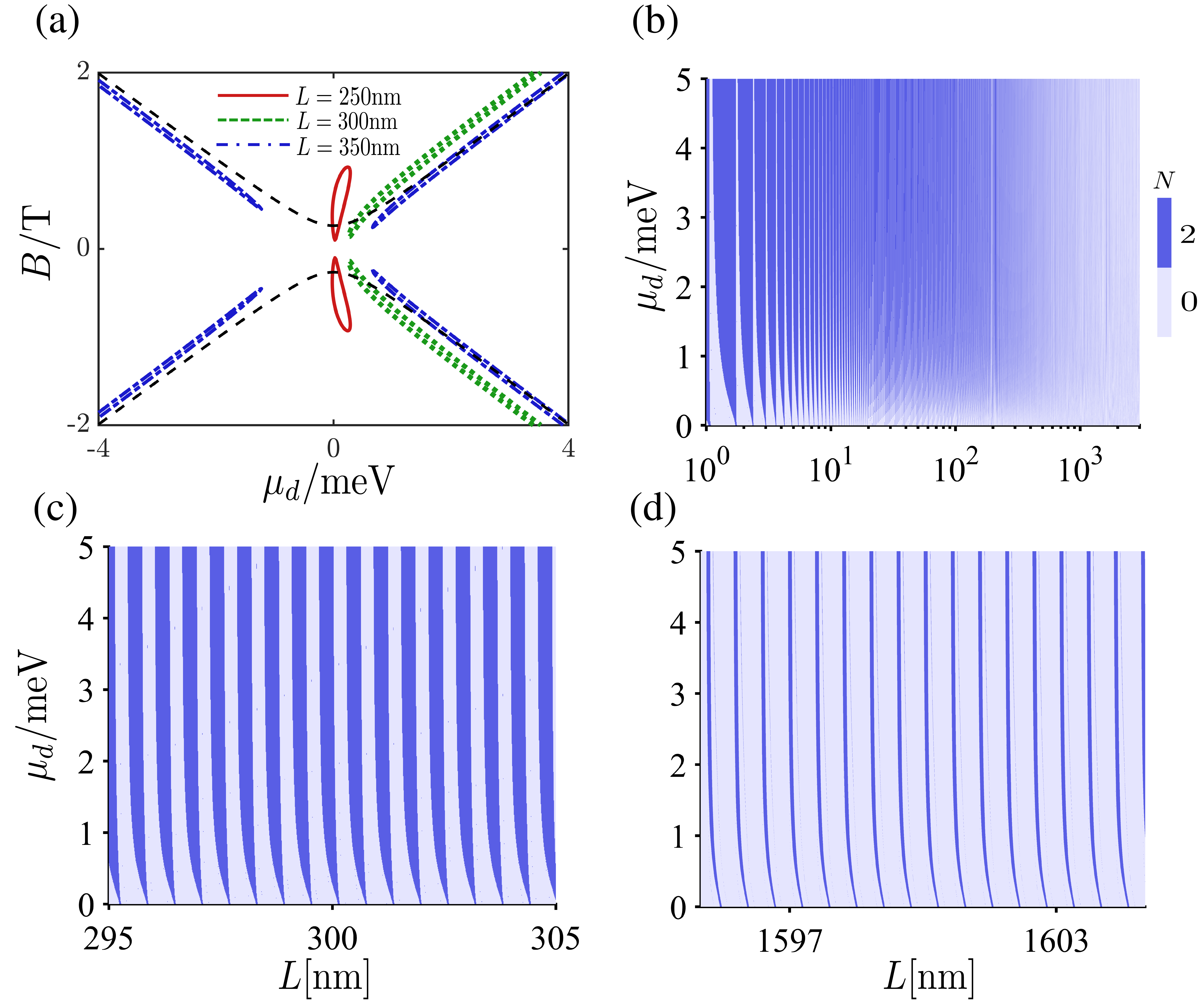}   \caption{(a) Condition for the existence of PMM in the practical experimental parameters. (b) Dependence of the number of PMMs on the length of SC $L$. Rapid oscillations in the number of PMMs as the SC length is varied in two representative regimes: the short-length range $295\text{--}305 \mathrm{nm}$ in (c), and the range near the superconducting coherence length, $1595\text{--}1605 \mathrm{nm}$ in (d). The parameters are given $g=35,\;\Delta_s \approx 0.34{\rm{meV}}, \mu_s = 1eV$, $t_s = 10 \mu_s$, $\xi_0\approx1600{\rm{nm}}$, $T \approx 32.6{\rm{meV}}$, $\alpha = 0.9 T$ and $\mu_d = 2\rm{meV}$}. 
    \label{phase diagramEx}
\end{figure}

In the practical QD–SC hybrid system~\cite{dvir2023realization, van2026single}, the QDs are formed in $\rm{InSb}$ nanowires, while the SC consists of an $\rm{Al/Pt}$ segment with a length of about $300 \rm{nm}$. For these material parameters, with a Landé factor of $g=35$, superconducting gap $\Delta_s\approx0.34 \rm{meV}$, chemical potential $\mu_s=1 \rm{eV}$, and tunneling strength $t_s=10\mu_s$, PMMs appear when the chemical potential of QD is tuned from $0.3$ to $1.6 \rm{meV}$ and the magnetic field is varied from $0.14$ to $1 \rm{T}$, as shown in Fig.~\ref{phase diagramEx}(a). Since the SC-induced couplings oscillate rapidly with the length of SC, with a period around $ 1\, \text{\AA}$, the existence condition for PMMs also depends sensitively on the SC length. As a result, the parameter region where PMMs emerge varies significantly for different lengths of SC [from $250 \rm{nm}$ (red solid line) to $350 \rm{nm}$ (blue dash-dotted line)], as illustrated in Fig.~\ref{phase diagramEx}(a). As the SC length increases, this length dependence gradually weakens, and the existence condition of PMM approaches that in the $L\to\infty$ limit (black dashed line in Fig.~\ref{phase diagramEx}(a)). Below, we turn to discuss the number of PMMs in the system, since it directly determines the zero-bias differential-conductance signal observed in experiments.

According to Eq. \eqref{eqofQD}, the number of PMMs in this hybrid system is determined by the dimension of the solution space of the corresponding linear equations~\cite{qiao2025size}, namely, the number of variables minus the rank of the coefficient matrix. Therefore, the number of PMMs is given by
\begin{equation}
    N = 2 \left\{4 - {\rm{rank\left[\mathbf{H}_d - (\mathbf{T} + \boldsymbol{\alpha})\cdot \mathbf{H}_s^{-1} \cdot(\mathbf{T} - \boldsymbol{\alpha})^\dagger\right]}}\right\}.
\end{equation}
It is shown that in the limit of an infinitely long SC, the SC-induced pairing strength, hopping strength, effective spin-flip coupling, and chemical-potential shift all simultaneously vanish. In this case, when the system parameters satisfy Eq.~\eqref{Ltoinfty}, the rank of the coefficient matrix is reduced to 2, so that the system supports four Majorana modes. This is consistent with our previous result~\cite{zhang2025poor}. Moreover, for a finite-length SC, the system hosts two PMMs only when the parameters satisfy Eq.~\eqref{sweet spot}; otherwise, no PMM exists. Because the coefficients in Eq.~\eqref{sweet spot} depend sensitively on the superconducting length through finite-length effects, the number of PMMs also depends sensitively on the superconducting length $L$, as shown in Fig.~\ref{phase diagramEx}(b). The Fig.~\ref{phase diagramEx} (c) and (d) provide enlarged views of some representative regions: the former shows the rapid oscillation (about $1 \AA$) of the PMM number at short length of SCs (around $L\approx300 \mathrm{nm}$), while the latter highlights the long-length regime (around coherence length of SC $\sim1600 \rm{ n m}$). Therefore, choosing an appropriate length of SC is favorable for experimentally accessing the PMM regime. We have determine the existence condition for PMMs and analyzed how their number oscillates with the superconducting length, but their spatial distribution in the full hybrid system has not yet been addressed.

\section{Conditions for PMMs Localized at the two ends of the hybrid system\label{sec: Spatial Separation}}
In this section, we examine the distribution of the two PMMs to determine whether they are localized separately at the two ends of the hybrid system. To further quantify their spatial distribution in the whole system, one must take into account their wave functions in both the QDs and the SC~\cite{zhang2025poor}. We find that the condition for PMMs to be localized at opposite ends of the system is governed by nonlinear equations involving the superconducting length, the QD chemical potential, and the Zeeman field. It is shown that these equations admit no exact solution. Therefore, in the QD–SC–QD hybrid system, strictly localized PMMs at the two opposite ends do not exist. We then consider the strong magnetic-field limit, where the condition admits approximate solutions. In this limit, the spin is fully polarized by the external magnetic field, and the PMM wave functions become localized near the two opposite ends of the hybrid system.  It is precisely in this strong magnetic-field limit that the ideal model~\cite{Flensberg2012} is approximately obtained from the practical system [see Eq. \eqref{flensberg_model}], which allows us to identify the corresponding generalized ``sweet spot'' in the realistic system.

The wave functions of the two PMMs are defined as
\begin{equation}
\begin{aligned}
    \psi_{r} &= [\mathbf{u}_{d,1}^{r},\;\mathbf{u}_{s}^{r},\;\mathbf{u}_{d,2}^{r}]^T,\\
    \psi_{i} &= [\mathbf{u}_{d,1}^{i},\;\mathbf{u}_{s}^{i},\;\mathbf{u}_{d,2}^{i}]^T.
    \end{aligned} \label{wavefunction_PMM}
\end{equation} 
Here, $\mathbf{u}_{d,1}^{r}$ and $\mathbf{u}_{d,2}^{r}$ represent the component of the wavefunction in the first and second QD, and $\mathbf{u}_{s}^{r}$ is the wavefunction of PMM in SC. By solving Eq.~\eqref{eqofQD}, the spatial distribution of the PMM wave function on two QDs are obtained as 
\begin{equation}
    \begin{aligned}
        \mathbf{u}_d^r \equiv \begin{bmatrix}
            \mathbf{u}_{d,1}^{r}\\
            \mathbf{u}_{d,2}^{r}
        \end{bmatrix} = \begin{bmatrix}
            (\bar{\mu}_d+h_d) \sigma_1 + (t - \Delta) \sigma_2\\
            \tilde{\mu}_d - \tilde{h}_d\\
            (t + \Delta) \sigma_1 + (\bar{\mu}_d+h_d) \sigma_2\\
            \tilde{t} + \tilde{\Delta}
        \end{bmatrix}. 
    \end{aligned}
\end{equation}
where $\sigma_1 = [\bar{\Delta}_s (\tilde{\mu}_d - \tilde{h}_d) + (\bar{\alpha} + \Delta_{sp})(\tilde{t} + \tilde{\Delta})]/Z$ and $\sigma_2 = [\bar{\Delta}_s (\tilde{t} + \tilde{\Delta}) - (\bar{\alpha} - \Delta_{sp}) (\tilde{\mu}_d - \tilde{h}_d)]/Z$. Here, $\tilde{\mu}_d$, $\tilde{h}_d$, $\tilde{t}$, and $\tilde{\Delta}$ are exactly the effective parameters appearing in the low-energy Hamiltonian~\eqref{practical_flensberg} of the spin-down sector. In addition, $\bar{\Delta}_s$ denotes the induced local pairing gap, while $\bar{\alpha}$ and $\Delta_{sp}$ represent, respectively, the SC-induced effective spin-flip coupling and the nonlocal pairing between opposite spins in the two QDs. These SC-induced couplings are defined in Eqs.~\eqref{eq_epsilon_deltabar}-\eqref{eq_delta_sp_delta}. It is straightforward to verify that the wave functions in the superconducting segment are obtained by
\begin{equation}
    \mathbf{u}_s^r = -\mathbf{H}_s^{-1} \cdot (\mathbf{T} - \boldsymbol{\alpha})^\dagger \cdot \mathbf{u}_d^r. \label{relation_of_SC_QD}
\end{equation} 
Therefore, when the parameters of the hybrid system satisfy Eq.~\eqref{sweet spot}, Eqs.~\eqref{eqofQD} and \eqref{relation_of_SC_QD} determine the spatial distribution of the PMM wave function. 

If this PMM is localized at the right end of the hybrid system, i.e., on the side of QD-2, then a necessary condition is that its wave-function weight vanishes on QD-1 while remaining finite on QD-2, namely, $\mathbf{u}_{d,1}^r=0$ and $\mathbf{u}_{d,2}^r\neq 0$. Through Eq.~\eqref{sweet spot} and $\mathbf{u}_{d,1}^r=0$, one finds that
\begin{equation}
\begin{aligned}
        &\tilde{\mu}_d - \tilde{h}_d = 0,\;\tilde{t} - \tilde{\Delta} = 0,\\
        &(\bar{\mu}_d+h_d) \sigma_1 + (t - \Delta) \sigma_2 = 0,
\end{aligned}\label{full_polarization3}
\end{equation}
and the wavefunction of PMM on the second QD becomes
\begin{equation}
\mathbf{u}_{d,2}^r = [(t + \Delta) \sigma_1 + (\bar{\mu}_d+h_d) \sigma_2,\;\tilde{t} + \tilde{\Delta}]^T.
\label{eq_second_QD}
\end{equation}
It should be promised that $\tilde{t} + \tilde{\Delta} \neq 0$ to keep the wavefunction of $\psi_r$ in QD-2 is nonzero. Moreover, it can also be shown that, once this condition \eqref{full_polarization3} is satisfied, the other PMM with wavefunction $\psi_i$ also localized at the left end of the hybrid system, namely, $\psi_i$ vanishes on the QD-2 while staying finite on QD-1. Therefore, Eq.~\eqref{full_polarization3} gives the condition for the existence of PMMs localized at the two ends of the hybrid system. In fact, Eq.~\eqref{full_polarization3} defines nonlinear equations for $(\mu_d,h_d,L)$, together with the constraints $Z\neq 0$ and $\tilde{t}+\tilde{\Delta}\neq 0$. As demonstrated in Appendix~\ref{appendixB}, under these constraints, the equations admit no solution when the superconducting length is finite. This means that, for a finite-length SC, PMMs that are strictly localized at the two opposite ends of the hybrid system cannot be realized. Therefore, the wave functions of the two PMMs have spatial overlap, but they remain zero modes. In the following, we examine whether Eq.~\eqref{full_polarization3} admits solutions in the limits of an infinitely long SC or an infinitely large Zeeman field.

If the length of SC is infinite  $L\to\infty$, the condition \eqref{full_polarization3} reduces to  Eq.~\eqref{Ltoinfty}. This indicates that, when the SC length approaches infinity, any PMMs supported by the system are automatically localized at the two opposite ends, in agreement with our previous result~\cite{zhang2025poor}. Moreover, it is shown that if the magnetic field is strong enough $h_d \gg \max\{T,\alpha\}$ so that $Z \to 0$, the third equation in Eq. \eqref{full_polarization3} can be satisfied approximately. In this case, the condition for the existence of PMMs localized at the two ends of the hybrid system reduces to
\begin{equation}
\bar{\mu}_d = h_d,\;t(L) = \Delta(L),
\end{equation}
where $t$ and $\Delta$ are the hopping and pairing strength directly induced by SC. Under this condition, Eq. \eqref{eq_second_QD} shows that $\mathbf{u}_{d,2}^r = [0,\;t+\Delta]^T$, indicating that $\mathbf{u}_{d,2}^r$ has no spin-up component and is fully polarized in the spin-down channel. In this limit, the corresponding low-energy effective Hamiltonian Eq. \eqref{practical_flensberg} reduced to
\begin{equation}
    H_{\downarrow,\rm{eff}} = \sum_{i=1,2} (\bar{\mu}_d - h_d) d_{i\downarrow}^\dagger d_{i\downarrow} - t d_{1\downarrow}^\dagger d_{2\downarrow} - \Delta d_{1\downarrow}^\dagger d_{2\downarrow}^\dagger + \rm{h.c.}. \label{flensberg_model}
\end{equation}
Therefore, the ideal model~\cite{Flensberg2012} is recovered from the realistic microscopic model. Moreover, the condition $\bar{\mu}_d = h_d$ and $t(L) = \Delta(L)$ defines the general `sweet spot' of the realistic hybrid system. In contrast to the `sweet spot' obtained in earlier ideal models~\cite{Flensberg2012}, our result is derived from a microscopic description that explicitly includes the microscopic parameters of the system (length of SC, chemical potential of QD, and so on).

\section{Conclusion}

We have studied the finite-length effects of a superconductor (SC) on poor man’s Majorana mode (PMMs) in a quantum-dot (QD)-SC-QD hybrid system. We obtained the local and nonlocal couplings in the QDs induced by SC (SC-induced couplings) through the low-energy effective Hamiltonian, including the shift of chemical potential, induced local pairing gap, interdot hopping, nonlocal pairing, and effective spin-flip coupling. It is shown that these couplings exhibit rapid oscillations and decay with the superconducting length, with a period of approximately $1\,\text{\AA}$. Moreover, in the strong magnetic-field limit, we recover the ideal model and obtain its corresponding existence condition.

We analytically derive the existence condition for PMMs in the full hybrid system, and the existence condition is determined by the SC-induced couplings. This condition is valid for arbitrary SC length and over a broad range of magnetic fields and tunneling strengths as the dressed effect of SC has been considered~\cite{Qiao2024,zhang2025poor}. We also showed that it reduces to the known result in the long-SC limit~\cite{zhang2025poor}, while extending it to realistic finite-length SC. We find that the number of PMMs depends very sensitively on the superconducting length due to the oscillatory behavior of the SC-induced couplings. In the long-SC limit, where the SC-induced interdot couplings vanish, the system supports four PMMs~\cite{zhang2025poor}; for finite SC lengths, the number of PMMs oscillates between zero and two with a period set by the Fermi wavelength, consistent with the finite-size oscillations reported in other hybrid systems~\cite{yue2025finite, qiao2025size}. Therefore, this strong length dependence is important for interpreting experiments on PMMs.

Finally, we determine the spatial distribution of the PMMs in the full hybrid system and the condition for the existence of PMMs that are separately localized at the two opposite ends of the system. We show that, for a finite-length SC, PMMs strictly localized at the two opposite ends of the system cannot be realized. When the SC length approaches infinity, we show that all PMMs supported by the system are localized at the two opposite ends~\cite{zhang2025poor}. In the strong magnetic-field limit, approximately localized PMMs emerge, and the corresponding ``sweet spot'' is determined in the practical QD–SC–QD hybrid system. The present approach can also be extended to larger QD--SC arrays, which are currently being explored experimentally~\cite{bordin2025enhanced, bordin2026probing, ten2025observation}.

\textit{Note added}. Recently, we became aware of a related work~\cite{alvarado2026optimal} in which the superconducting segment is treated as a QD-SC hybrid region to investigate how the microscopic structure of the hybrid section modifies the effective couplings, sweet-spot conditions, and localization properties of Majorana modes.

\section*{acknowledgment \label{sec7}}

This work was supported by the Science Challenge Project (Grant No.TZ2025017), the National Natural Science Foundation of China (NSFC) (Grant No. 12088101, 12547124), and the China Postdoctoral Science Foundation (Grant No. 2025M784438).

\section*{data availability \label{sec8}}
The data that support the findings of this article are not publicly available. The data are available from the authors upon reasonable request.

\twocolumngrid
% \onecolumngrid
\appendix   %仅一个附录时用appendix，否则\appendices
\setcounter{table}{0}   %从0开始编号，显示出来表会A1开始编号
\setcounter{figure}{0}
%定义编号格式，在数字序号前加字符“A"
\renewcommand{\thetable}{A\arabic{table}}
\renewcommand{\thefigure}{A\arabic{figure}}

\section{Effective coupling induced by Superconductor and the Condition of the existence of PMM \label{appendixA}}

In this Appendix, we provide the analytical form of the effective couplings between quantum dots (QDs) induced by the superconductor (SC). For the two QDs coupled to the finite-length SC, its Hamiltonian [Eq. \eqref{H_SC_l}] can be rewritten as
\begin{equation}
    \begin{aligned}
        H = \frac12 \mathbf{C}^{\dagger} \cdot \mathbf{H} \cdot \mathbf{C},\quad  \mathbf{H}=\begin{bmatrix}\mathbf{h} & \mathbf{p}\\
\mathbf{p^{\dagger}} & -\mathbf{h}
\end{bmatrix}.
\end{aligned}
\end{equation} 
Here, the vector operator is $\mathbf{C}^{\dagger} = [\mathbf{d}^{\dagger}, \mathbf{c}^{\dagger}, \mathbf{d}, \mathbf{c}]$ with $\mathbf{d}^\dagger = [d_{1\uparrow}^\dagger, d_{1\downarrow}, d_{2 \uparrow}^\dagger, d_{2\downarrow}]^T$ and $\mathbf{c}=[c_{1\uparrow}^\dagger,c_{1\downarrow},\ldots,c_{N\uparrow}^\dagger,c_{N\downarrow}]^{T}$, $\mathbf{h}$ and $\mathbf{p}$ are $(2N+4)\times(2N+4)$ matrices
\begin{equation}
    \mathbf{h}=\begin{bmatrix}\mathbf{H}_{d} & \mathbf{T}\\
\mathbf{T}^{\dagger} & \mathbf{H}_{s}
\end{bmatrix},\quad\mathbf{p}=\begin{bmatrix}\mathbf{0} & \boldsymbol{\alpha}\\
- \boldsymbol{\alpha}^\dagger & \mathbf{0}
\end{bmatrix},
\end{equation}
where the Hamiltonian matrices of the QDs, SC without the pairing term, and the tunneling term between the QDs and SC are
\begin{equation}
\begin{aligned}
\mathbf{H}_d &= (\mu_d \sigma_z + h_d \sigma_0) \otimes \sigma_0\\
\mathbf{T} &= \begin{bmatrix}T_1 & 0 &\dots & T_N &0\\
0 & -T_1 & \dots & 0 & -T_N\\
\tilde{T}_1 & 0 & \dots & \tilde{T}_{N} & 0\\
 0 &-\tilde{T}_1 & \dots &0 & - \tilde{T}_{N} 
\end{bmatrix}_{4\times 2N},\\
\boldsymbol{\alpha} &= \begin{bmatrix}0 & \alpha_1 &\dots & 0 &\alpha_N\\
\alpha_1 & 0 & \dots & \alpha_N & 0\\
0 & -\tilde{\alpha}_1 & \dots & 0 & -\tilde{\alpha}_N\\
 -\tilde{\alpha}_1 &0 & \dots &-\tilde{\alpha}_N & 0 
\end{bmatrix}_{4\times 2N},\\
\mathbf{H}_{s} &= {\rm{diag}}\{H_{sc}(i=1),\dots,H_{sc}(i=N)\},\\
H_{sc}(i) &= (\bar{\mu}_{s}(i) \sigma_{z} - \Delta_s \sigma_{x})\otimes\tau_{0}.
\end{aligned} \label{Hd_T_alpha_Hs}
\end{equation}
Here, $\sigma_0 = {\rm{diag}}(1,1)$ is identity matrix, $\sigma_{x,y,z}$ are the Pauli matrices, $\tau_0 =\mathbb{I}_{N \times N}$ is the $N \times N$ identity, and $\bar{\mu}_{s}(l) = t_s\{\cos [\pi l/(N+1)] - \cos [\pi /(N+1)]\} + \mu_s$ is the dispersion relation of $H_s$ except for pairing term. $o_l = o \sqrt{2/(N+1)}\sin[\pi l /(N+1)] = (-1)^{l+1} \tilde{o}_l$ with $o = T,\alpha$ are the strength of tunneling ($o=T$) and spin-flipping coupling ($o=\alpha$) between QDs and mode $l$ of SC. 

The eigen-equation of the Hamiltonian is: $\mathbf{H} \psi_\nu = E_{\nu} \psi_\nu$ where $\psi_\nu = [\mathbf{u}_\nu\;\mathbf{v}_{\nu}]^T$ with $\mathbf{u}_\nu = [\mathbf{u}_\nu^d\;\mathbf{u}_\nu^s]^T$ and $\mathbf{v}_\nu = [\mathbf{v}_\nu^d\;\mathbf{v}_\nu^s]^T$. Due to the definition of Majorana fermion with $\gamma_\nu^{\dagger} = \gamma_{\nu}$, it is shown that $\mathbf{u}_\nu = \mathbf{v}_\nu^*$.  Furthermore, the particle-hole symmetry inherent in the hybrid system imposes $E_\nu = 0$. The eigen-equation of $E_\nu = 0$ is
\begin{equation}
    \begin{aligned}
        (\mathbf{h}+\mathbf{p})\cdot\mathbf{u}^r = 0,\;\;(\mathbf{h}-\mathbf{p})\cdot\mathbf{u}^i = 0  
    \end{aligned}\label{equdus}
\end{equation}
where $\mathbf{u}^{r(s)}$ are the real (imaginary) part of $\mathbf{u}$ and the subscript $\nu$ has been omitted. In this case,$\mathbf{u}^r = [\mathbf{u}_d^r\;\mathbf{u}_s^r]^T$ and the corresponding equation of $\mathbf{u}_d^r$ is $[\mathbf{H}_d - (\mathbf{T} + \boldsymbol{\alpha})\cdot \mathbf{H}_s^{-1}\cdot (\mathbf{T} - \boldsymbol{\alpha})] \cdot \mathbf{u}_d^r \equiv \mathbf{B} \cdot \mathbf{u}_d^r= 0$ where
\begin{equation}
\begin{aligned}
    \mathbf{B} &= [-\bar{\Delta}_s \sigma_x + (\mu_d - \bar{\epsilon}) \sigma_z + h_d \sigma_0] \otimes \sigma_0 \\
    &+ i (\Delta \sigma_z - \bar{\alpha} \sigma_x) \otimes \sigma_y - (t \sigma_z + \Delta_{sp} \sigma_x)\otimes\sigma_x.
\end{aligned}
\end{equation}
Here, $\bar{\Delta}_s,\; \bar{\epsilon},\;t,\;\Delta,\;\bar{\alpha},\;\Delta_{sp}$ represent the coupling induced by the SC. Among them, $\bar{\Delta}_s$ and $\bar{\epsilon}$ are the pairing strength and shift of chemical potential of each QD, respectively. And $t,\;\bar{\alpha}$ are the strength of hopping and spin-flippingal coupling; $\Delta,\;\Delta_{sp}$ are the pairing strength between the two QDs. Specifically,
\begin{equation}
\begin{aligned}
    \bar{\epsilon} &= \sum_{i=1}^N \frac{\abs{T_i}^2 + \abs{\alpha_i}^2}{\epsilon_i^2 + \Delta_s^2} \epsilon_i,\;\bar{\Delta}_s = \sum_{i=1}^N \frac{\abs{T_i}^2 + \abs{\alpha_i}^2}{\epsilon_i^2 + \Delta_s^2} \Delta_s,\\
    t &= \sum_{i=1}^N \frac{T_i \tilde{T}_i - \alpha_i \tilde{\alpha}_i}{\epsilon_i^2 + \Delta_s^2} \epsilon_i,\;\Delta_{sp} = \sum_{i=1}^N \frac{T_i \tilde{T}_i - \alpha_i \tilde{\alpha}_i}{\epsilon_i^2 + \Delta_s^2} \Delta_s,\\
    \bar{\alpha} &= \sum_{i=1}^N \frac{\alpha_i \tilde{T}_i + T_i \tilde{\alpha}_i}{\epsilon_i^2 + \Delta_s^2} \epsilon_i,\;\Delta = \sum_{i=1}^N \frac{\alpha_i \tilde{T}_i + T_i \tilde{\alpha}_i}{\epsilon_i^2 + \Delta_s^2} \Delta_s.
\end{aligned}
\end{equation}
We further present the dependence of these superconductivity-induced couplings on the length of SC. For example, it is shown that the shift of chemical potential is
\begin{equation}
    \begin{aligned}
        \bar{\epsilon} &= \sum_{i=1}^N \frac{\abs{T_i}^2 + \abs{\alpha_i}^2}{\epsilon_i^2 + \Delta_s^2} \epsilon_i = \sum_{i=1}^N (T_i^2 + \alpha_i^2) {\rm{Re}} \left(\frac{1}{\epsilon_i - i\Delta_s}\right)\\
        &\approx \frac{(T^2 + \alpha^2)a}{L} \sum_{n=-\infty}^{\infty} {\rm{Re}} \left(\frac{1}{\bar{\mu}_s - i \Delta_s - \frac{\hbar^2 \pi^2}{2 m_s L^2}n^2}\right)\\
        &\approx \Delta_0  \frac{\sin( 2 k_F L)}{\cosh\left(\frac{L}{\xi_0}\right) - \cos(2k_F L)}
    \end{aligned}\label{shift_mu}
\end{equation}
where $\Delta_0 = (T^2 + \alpha^2)a/(\Delta_s \xi_0)$. Here, $\bar{\mu}_s = \mu_s + \hbar^2 \pi^2/(2m_sL)$ is the effective chemical of SC, $m_s$ is the mass of electron, $L$ is the length of SC, $k_F = \sqrt{2m_s\bar{\mu}_s}/\hbar$ and $\xi_0 = \hbar v_F/(2\Delta_s)$ are the effective Fermi velocity and coherence length of SC. Moreover, in deriving Eq. \eqref{shift_mu}, we have also employed the wide-band approximation ($t_s\gg\Delta_s$) and assumed that the superconducting gap provides the lowest energy scale in the system ($t_s\gg\mu_s\gg\Delta_s$). Similarly, the length dependence of the remaining five SC-induced couplings can be derived in the same manner
\begin{equation}
    \begin{aligned}
        \bar{\Delta}_s &= \Delta_0 \frac{\sinh\left(\frac{L}{\xi_0}\right)}{\cosh\left(\frac{L}{\xi_0}\right) - \cos(2k_F L)},\\
    \end{aligned}\label{barDeltas}
\end{equation}
and $\Delta = \Delta_1 g(L),\;\Delta_{sp} = \Delta_2 g(L),\;\bar{\alpha} = \Delta_1 h(L),\;t = \Delta_2 h(L)$ with $\Delta_1 = - 4 T\alpha a/(\Delta_s \xi_0)$ and $\Delta_2 = - 2 (T^2 - \alpha^2) a/(\Delta_s \xi_0)$ and
\begin{equation}
    \begin{aligned}
        g(L) &= \frac{\sinh\left(\frac{L}{2\xi_0}\right) \cos(k_F L)}{\cosh\left(\frac{L}{\xi_0}\right) - \cos(2k_F L)},\\
        h(L) &= \frac{\cosh\left(\frac{L}{2\xi_0}\right) \sin(k_F L)}{\cosh\left(\frac{L}{\xi_0}\right) - \cos(2k_F L)}.
    \end{aligned} \label{Delta_alpha}
\end{equation}
And the relation $\sum_{i=-\infty}^\infty (-1)^{i+1} / (\chi_1 i^2 - \chi_2) = \pi/\sqrt{\chi_1 \chi_2} \csc(\pi \sqrt{\chi_2/\chi_1})$ has been used. It is found that when $L \gg \xi_0$, the coupling can be approximated as
\begin{equation}
\begin{aligned}
    \bar{\epsilon} &= 2 \Delta \sin(2 k_F L) e^{- \frac{L}{\xi_0}},\;\bar{\Delta}_s = \Delta f(L),\\
    \Delta &= \Delta_1 f(L) \cos(k_F L) e^{- \frac{L}{2 \xi_0}},\;\bar{\alpha} = \Delta_1 f(L) \sin(k_F L) e^{- \frac{L}{2\xi_0}},\\
    \Delta_{sp} &= \Delta_2 f(L) \cos(k_F L) e^{- \frac{L}{2 \xi_0}},\;t = \Delta_2 f(L) \sin(k_F L) e^{- \frac{L}{2 \xi_0}}
    \end{aligned}
\end{equation}
with $f(L) = 1 + 2 \cos(2 k_F L) \exp(-L/\xi_0)$. Apparently, when $L \to \infty$, only the $\bar{\Delta}_s$ is nonzero. 

Furthermore, the condition of the existence of PMM is ${\rm Det}(\mathbf{B}) = 0$ which is
\begin{equation}
    \left[\tilde{\mu}_d(L) - \tilde{h}_d(L) \right]^2 - \tilde{t}(L)^2 + \tilde{\Delta}(L)^2 = 0 \label{sweet_spot_appendix}
\end{equation}
where $\tilde{\mu}_d(L),\tilde{h}_d(L),\tilde{t}(L),\tilde{\Delta}(L)$ are defined in the main text. It can be demonstrated that when the parameters of the hybrid system satisfy Eq. \eqref{sweet_spot_appendix}, the equation of $\mathbf{u}_d^i$ also has a solution.

\section{Proof of the absence of PMMs localized at the two ends of hybrid system \label{appendixB}}
In this Appendix, we will show that there are no PMMs localized at the two ends of the hybrid system. The condition of PMMs localized at the two ends of the system is shown by
\begin{align}
\tilde{\mu}_d - \tilde{h}_d &= 0, \label{eq1}\\
\qquad \tilde{t} - \tilde{\Delta} &= 0, \label{eq2}\\
(\bar{\mu}_d + h_d)\sigma_1 + (t - \Delta)\sigma_2 &= 0,\label{eq3}
\end{align}
with constraints $\tilde{t} + \tilde{\Delta} \neq 0$ and $Z \equiv (\bar{\mu}_d + h_d)^2 + \Delta^2 - t^2 \neq 0$. Here, $\sigma_1 = [\bar{\Delta}_s (\tilde{\mu}_d - \tilde{h}_d) + (\bar{\alpha} + \Delta_{sp})(\tilde{t} + \tilde{\Delta})]/Z$ and $\sigma_2 = [\bar{\Delta}_s (\tilde{t} + \tilde{\Delta}) - (\bar{\alpha} - \Delta_{sp}) (\tilde{\mu}_d - \tilde{h}_d)]/Z$. This condition is Eq.~\eqref{full_polarization3} in the main text. We will demonstrate that there is no solution of Eq.~\eqref{full_polarization3} with the above constraints. First, using Eq. \eqref{eq1} and the definition of $\sigma_1$ and $\sigma_2$, Eq. \eqref{eq3} shows that
\begin{equation}
(\bar{\alpha} + \Delta_{sp})(\bar{\mu}_d + h_d) + \bar{\Delta}_s(t - \Delta) = 0.\label{eq_of_3}
\end{equation}
Equation~\eqref{eq_of_3} already implies that $\bar{\alpha}+\Delta_{sp}$ and $t-\Delta$ cannot vanish simultaneously. Indeed, if $\bar{\alpha}+\Delta_{sp}=0$, then Eq.~\eqref{eq_of_3} together with $\bar{\Delta}_s\neq 0$ would require $t-\Delta=0$; however, according to the definitions of $\bar{\alpha}$, $\Delta_{sp}$, $t$, and $\Delta$ given in Appendix~\ref{appendixA}, these two equalities cannot hold at the same time. Moreover, if $\bar{\mu}_d + h_d = 0$ then $t - \Delta$ must be vanished as $\bar{\Delta}_s \neq 0$. This means $Z = 0$ and it is contradictory to the constraint $Z \neq 0$. Therefore, it is shown that $\bar{\mu}_d + h_d \neq 0$. Next, Eq. \eqref{eq2} is simplified that
\begin{equation}
(t - \Delta)(Z - \bar{\Delta}_s^2) - (\bar{\alpha} + \Delta_{sp})^2(t + \Delta) - 2 \bar{\Delta}_s (\bar{\mu}_d + h_d)(\Delta_{sp} + \bar{\alpha}) = 0.\label{eq_of_2}
\end{equation}
Through Eq. \eqref{eq_of_3} and \eqref{eq_of_2}, it is found that
\begin{equation}
\left[(\bar{\mu}_d + h_d)^2 + \bar{\Delta}_s^2\right] Z = 0.
\end{equation}
Since $(\bar{\mu}_d + h_d)^2 + \bar{\Delta}_s^2 > 0$, it follows that
\begin{equation}
Z = 0.
\end{equation}
As $Z \neq 0$ is the constraint of the condition~\eqref{full_polarization3}, the condition has no solution. Therefore, there are no PMMs localized at the two ends of the hybrid system.

\twocolumngrid
% \onecolumngrid
\bibliography{main}

@article{kitaev2001unpaired,
  title={Unpaired Majorana fermions in quantum wires},
  author={Kitaev, A Yu},
  journal={Phys. Usp.},
  volume={44},
  number={10S},
  pages={131},
  year={2001},
  publisher={IOP Publishing},
url = {https://iopscience.iop.org/article/10.1070/1063-7869/44/10S/S29}
}

@article{PhysRevLett.104.040502,
  title = {Generic New Platform for Topological Quantum Computation Using Semiconductor Heterostructures},
  author = {Sau, Jay D. and Lutchyn, Roman M. and Tewari, Sumanta and Das Sarma, S.},
  journal = {Phys. Rev. Lett.},
  volume = {104},
  issue = {4},
  pages = {040502},
  numpages = {4},
  year = {2010},
  month = {Jan},
  publisher = {American Physical Society},
  doi = {10.1103/PhysRevLett.104.040502},
  url = {https://link.aps.org/doi/10.1103/PhysRevLett.104.040502}
}

@article{alicea2012new,
  title={New directions in the pursuit of Majorana fermions in solid state systems},
  author={Alicea, Jason},
  journal={Rep. Prog. Phys.},
  volume={75},
  number={7},
  pages={076501},
  year={2012},
  publisher={IOP Publishing},
url = {https://iopscience.iop.org/article/10.1088/0034-4885/75/7/076501}
}

@article{lutchyn2018majorana,
  title={Majorana zero modes in superconductor--semiconductor heterostructures},
  author={Lutchyn, Roman M and Bakkers, Erik PAM and Kouwenhoven, Leo P and Krogstrup, Peter and Marcus, Charles M and Oreg, Yuval},
  journal={Nat. Rev. Mater.},
  volume={3},
  number={5},
  pages={52--68},
  year={2018},
  publisher={Nature Publishing Group UK London},
url = {https://www.nature.com/articles/s41578-018-0003-1#citeas}
}

@article{prada2020andreev,
  title={From Andreev to Majorana bound states in hybrid superconductor--semiconductor nanowires},
  author={Prada, Elsa and San-Jose, Pablo and de Moor, Michiel WA and Geresdi, Attila and Lee, Eduardo JH and Klinovaja, Jelena and Loss, Daniel and Nyg{\aa}rd, Jesper and Aguado, Ram{\'o}n and Kouwenhoven, Leo P},
  journal={Nat. Rev. Phys.},
  volume={2},
  number={10},
  pages={575--594},
  year={2020},
  publisher={Nature Publishing Group UK London},
url = {https://www.nature.com/articles/s42254-020-0228-y}
}

@article{PhysRevLett.105.077001,
  title = {Majorana Fermions and a Topological Phase Transition in Semiconductor-Superconductor Heterostructures},
  author = {Lutchyn, Roman M. and Sau, Jay D. and Das Sarma, S.},
  journal = {Phys. Rev. Lett.},
  volume = {105},
  issue = {7},
  pages = {077001},
  numpages = {4},
  year = {2010},
  month = {Aug},
  publisher = {American Physical Society},
  doi = {10.1103/PhysRevLett.105.077001},
  url = {https://link.aps.org/doi/10.1103/PhysRevLett.105.077001}
}

@article{PhysRevLett.105.177002,
  title = {Helical Liquids and Majorana Bound States in Quantum Wires},
  author = {Oreg, Yuval and Refael, Gil and von Oppen, Felix},
  journal = {Phys. Rev. Lett.},
  volume = {105},
  issue = {17},
  pages = {177002},
  numpages = {4},
  year = {2010},
  month = {Oct},
  publisher = {American Physical Society},
  doi = {10.1103/PhysRevLett.105.177002},
  url = {https://link.aps.org/doi/10.1103/PhysRevLett.105.177002}
}

@article{PhysRevLett.100.096407,
  title = {Superconducting Proximity Effect and Majorana Fermions at the Surface of a Topological Insulator},
  author = {Fu, Liang and Kane, C. L.},
  journal = {Phys. Rev. Lett.},
  volume = {100},
  issue = {9},
  pages = {096407},
  numpages = {4},
  year = {2008},
  month = {Mar},
  publisher = {American Physical Society},
  doi = {10.1103/PhysRevLett.100.096407},
  url = {https://link.aps.org/doi/10.1103/PhysRevLett.100.096407}
}

@article{PhysRevB.82.184516,
  title = {Chiral topological superconductor from the quantum Hall state},
  author = {Qi, Xiao-Liang and Hughes, Taylor L. and Zhang, Shou-Cheng},
  journal = {Phys. Rev. B},
  volume = {82},
  issue = {18},
  pages = {184516},
  numpages = {5},
  year = {2010},
  month = {Nov},
  publisher = {American Physical Society},
  doi = {10.1103/PhysRevB.82.184516},
  url = {https://link.aps.org/doi/10.1103/PhysRevB.82.184516}
}

@article{PhysRevB.96.075161,
  title = {Andreev bound states versus Majorana bound states in quantum dot-nanowire-superconductor hybrid structures: Trivial versus topological zero-bias conductance peaks},
  author = {Liu, Chun-Xiao and Sau, Jay D. and Stanescu, Tudor D. and Das Sarma, S.},
  journal = {Phys. Rev. B},
  volume = {96},
  issue = {7},
  pages = {075161},
  numpages = {29},
  year = {2017},
  month = {Aug},
  publisher = {American Physical Society},
  doi = {10.1103/PhysRevB.96.075161},
  url = {https://link.aps.org/doi/10.1103/PhysRevB.96.075161}
}

@article{PhysRevB.86.100503,
  title = {Near-zero-energy end states in topologically trivial spin-orbit coupled superconducting nanowires with a smooth confinement},
  author = {Kells, G. and Meidan, D. and Brouwer, P. W.},
  journal = {Phys. Rev. B},
  volume = {86},
  issue = {10},
  pages = {100503},
  numpages = {5},
  year = {2012},
  month = {Sep},
  publisher = {American Physical Society},
  doi = {10.1103/PhysRevB.86.100503},
  url = {https://link.aps.org/doi/10.1103/PhysRevB.86.100503}
}

@article{li2014probing,
  title={Probing zero modes of a defect in a Kitaev quantum wire},
  author={Li, Sheng-Wen and Li, Zeng-Zhao and Cai, CY and Sun, C. P.},
  journal={Phys. Rev. B},
  volume={89},
  number={13},
  pages={134505},
  year={2014},
url={https://journals.aps.org/prb/abstract/10.1103/PhysRevB.89.134505},
  publisher={APS}
}

@article{PhysRevX.13.031031,
  title = {Tunable Crossed Andreev Reflection and Elastic Cotunneling in Hybrid Nanowires},
  author = {Bordin, Alberto and Wang, Guanzhong and Liu, Chun-Xiao and ten Haaf, Sebastiaan L. D. and van Loo, Nick and Mazur, Grzegorz P. and Xu, Di and van Driel, David and Zatelli, Francesco and Gazibegovic, Sasa and Badawy, Ghada and Bakkers, Erik P. A. M. and Wimmer, Michael and Kouwenhoven, Leo P. and Dvir, Tom},
  journal = {Phys. Rev. X},
  volume = {13},
  issue = {3},
  pages = {031031},
  numpages = {9},
  year = {2023},
  month = {Sep},
  publisher = {American Physical Society},
  doi = {10.1103/PhysRevX.13.031031},
  url = {https://link.aps.org/doi/10.1103/PhysRevX.13.031031}
}

@article{luethi2024perfect,
  title={From perfect to imperfect poor man's Majoranas in minimal Kitaev chains},
  author={Luethi, Melina and Legg, Henry F and Loss, Daniel and Klinovaja, Jelena},
  journal={Phys. Rev. B},
  volume={110},
  number={24},
  pages={245412},
  year={2024},
  publisher={APS},
url = {https://journals.aps.org/prb/abstract/10.1103/PhysRevB.110.245412}
}

@article{Qiao2024,
  title = {Dressed Majorana Fermion in a Hybrid Nanowire},
  author = {Qiao, Guo-Jian and Yue, Xin and Sun, C. P.},
  journal = {Phys. Rev. Lett.},
  volume = {133},
  issue = {26},
  pages = {266605},
  numpages = {5},
  year = {2024},
  month = {Dec},
  publisher = {APS},
  url = {https://link.aps.org/doi/10.1103/PhysRevLett.133.266605}
}

@article{liu2024enhancing,
  title={Enhancing the excitation gap of a quantum-dot-based Kitaev chain},
  author={Liu, Chun-Xiao and Bozkurt, A Mert and Zatelli, Francesco and ten Haaf, Sebastiaan LD and Dvir, Tom and Wimmer, Michael},
  journal={Commun. Phys.},
  volume={7},
  number={1},
  pages={235},
  year={2024},
  publisher={Nature Publishing Group UK London},
url={https://www.nature.com/articles/s42005-024-01715-5}
}

@article{dvir2023realization,
  title={Realization of a minimal Kitaev chain in coupled quantum dots},
  author={Dvir, Tom and Wang, Guanzhong and van Loo, Nick and Liu, Chun-Xiao and Mazur, Grzegorz P and Bordin, Alberto and Ten Haaf, Sebastiaan LD and Wang, Ji-Yin and van Driel, David and Zatelli, Francesco and others},
  journal={Nat.},
  volume={614},
  number={7948},
  pages={445--450},
  year={2023},
  publisher={Nature Publishing Group UK London},
url={https://www.nature.com/articles/s41586-022-05585-1}
}

@article{ten2024two,
  title={A two-site Kitaev chain in a two-dimensional electron gas},
  author={Ten Haaf, Sebastiaan LD and Wang, Qingzhen and Bozkurt, A Mert and Liu, Chun-Xiao and Kulesh, Ivan and Kim, Philip and Xiao, Di and Thomas, Candice and Manfra, Michael J and Dvir, Tom and others},
  journal={Nat.},
  volume={630},
  number={8016},
  pages={329--334},
  year={2024},
  publisher={Nature Publishing Group UK London},
url={https://www.nature.com/articles/s41586-024-07434-9}
}

@article{zatelli2024robust,
  title={Robust poor man’s Majorana zero modes using Yu-Shiba-Rusinov states},
  author={Zatelli, Francesco and van Driel, David and Xu, Di and Wang, Guanzhong and Liu, Chun-Xiao and Bordin, Alberto and Roovers, Bart and Mazur, Grzegorz P and van Loo, Nick and Wolff, Jan C and others},
  journal={Nat. Commun.},
  volume={15},
  number={1},
  pages={7933},
  year={2024},
  publisher={Nature Publishing Group UK London},
url={https://www.nature.com/articles/s41467-024-52066-2}
}

@article{sau2012realizing,
  title={Realizing a robust practical Majorana chain in a quantum-dot-superconductor linear array},
  author={Sau, Jay D and Sarma, S Das},
  journal={Nat. Commun.},
  volume={3},
  number={1},
  pages={964},
  year={2012},
  publisher={Nature Publishing Group UK London},
url={https://www.nature.com/articles/s42005-024-01715-5}
}

@article{Qiao_2025,
doi = {10.1088/1572-9494/adc5e8},
url = {https://dx.doi.org/10.1088/1572-9494/adc5e8},
year = {2025},
month = {may},
publisher = {IOP Publishing},
volume = {77},
number = {9},
pages = {095103},
author = {Qiao, Guo-Jian and Zhang, Zhi-Lei and Li, Sheng-Wen and Sun, C. P.},
title = {Controlling a superconducting transistor by coherent light},
journal = {Communications in Theoretical Physics}
}

@article{onedimensionSC,
  title = {Finite-size effects in a nanowire strongly coupled to a thin superconducting shell},
  author = {Reeg, Christopher and Loss, Daniel and Klinovaja, Jelena},
  journal = {Phys. Rev. B},
  volume = {96},
  issue = {12},
  pages = {125426},
  numpages = {12},
  year = {2017},
  month = {Sep},
  publisher = {APS},
  url = {https://link.aps.org/doi/10.1103/PhysRevB.96.125426}
}

@article{fateofDPPM,
  title = {Fate of poor man's Majoranas in the long Kitaev chain limit},
  author = {Luethi, Melina and Legg, Henry F. and Loss, Daniel and Klinovaja, Jelena},
  journal = {Phys. Rev. B},
  volume = {111},
  issue = {11},
  pages = {115419},
  numpages = {20},
  year = {2025},
  month = {Mar},
  publisher = {American Physical Society},
  doi = {10.1103/PhysRevB.111.115419},
  url = {https://link.aps.org/doi/10.1103/PhysRevB.111.115419}
}

@article{Creating2022,
  title = {Creating and detecting poor man's Majorana bound states in interacting quantum dots},
  author = {Tsintzis, Athanasios and Souto, Rub\'en Seoane and Leijnse, Martin},
  journal = {Phys. Rev. B},
  volume = {106},
  issue = {20},
  pages = {L201404},
  numpages = {6},
  year = {2022},
  month = {Nov},
  publisher = {American Physical Society},
  doi = {10.1103/PhysRevB.106.L201404},
  url = {https://link.aps.org/doi/10.1103/PhysRevB.106.L201404}
}

@article{Flensberg2012,
  title = {Parity qubits and poor man's Majorana bound states in double quantum dots},
  author = {Leijnse, Martin and Flensberg, Karsten},
  journal = {Phys. Rev. B},
  volume = {86},
  issue = {13},
  pages = {134528},
  numpages = {7},
  year = {2012},
  month = {Oct},
  publisher = {American Physical Society},
  doi = {10.1103/PhysRevB.86.134528},
  url = {https://link.aps.org/doi/10.1103/PhysRevB.86.134528}
}

@article{Qiao_2022,
  title = {Magnetic field constraint for Majorana zero modes in a hybrid nanowire},
  author = {Qiao, Guo-Jian and Li, Sheng-Wen and Sun, C. P.},
  journal = {Phys. Rev. B},
  volume = {106},
  issue = {10},
  pages = {104517},
  numpages = {16},
  year = {2022},
  month = {Sep},
  publisher = {American Physical Society},
  doi = {10.1103/PhysRevB.106.104517},
  url = {https://link.aps.org/doi/10.1103/PhysRevB.106.104517}
}

@article{Sankar_Das_Sarma2023,
   author = {Sankar Das Sarma},
   doi = {10.1038/s41567-022-01900-9},
   issn = {1745-2481},
   issue = {2},
   journal = {Nature Physics},
   pages = {165-170},
   title = {In search of Majorana},
   volume = {19},
   year = {2023},
}

@article{Yue2023,
  title = {Refined Majorana phase diagram in a topological insulator--superconductor hybrid system},
  author = {Yue, Xin and Qiao, Guo-Jian and Sun, C. P.},
  journal = {Phys. Rev. B},
  volume = {108},
  issue = {19},
  pages = {195405},
  numpages = {6},
  year = {2023},
  month = {Nov},
  publisher = {American Physical Society},
  doi = {10.1103/PhysRevB.108.195405},
  url = {https://link.aps.org/doi/10.1103/PhysRevB.108.195405}
}

@article{flensberg2021engineered,
  title={Engineered platforms for topological superconductivity and Majorana zero modes},
  author={Flensberg, Karsten and von Oppen, Felix and Stern, Ady},
  journal={Nat. Rev. Mater.},
  volume={6},
  number={10},
  pages={944--958},
  year={2021},
  publisher={Nature Publishing Group UK London},
url = {https://www.nature.com/articles/s41578-021-00336-6}
}

@article{zhang2025poor,
  title={Poor Man's Majoranon in Two Quantum Dots Dressed by Superconducting Quasi-Excitations},
  author={Zhang, Zhi-Lei and Qiao, Guo-Jian and Sun, CP},
  journal={arXiv preprint arXiv:2506.10367},
  year={2025}
}

@article{PhysRevLett.111.060501,
  title = {Coupling Spin Qubits via Superconductors},
  author = {Leijnse, Martin and Flensberg, Karsten},
  journal = {Phys. Rev. Lett.},
  volume = {111},
  issue = {6},
  pages = {060501},
  numpages = {5},
  year = {2013},
  month = {Aug},
  publisher = {American Physical Society},
  doi = {10.1103/PhysRevLett.111.060501},
  url = {https://link.aps.org/doi/10.1103/PhysRevLett.111.060501}
}

@article{PhysRevLett.129.267701,
  title = {Tunable Superconducting Coupling of Quantum Dots via Andreev Bound States in Semiconductor-Superconductor Nanowires},
  author = {Liu, Chun-Xiao and Wang, Guanzhong and Dvir, Tom and Wimmer, Michael},
  journal = {Phys. Rev. Lett.},
  volume = {129},
  issue = {26},
  pages = {267701},
  numpages = {8},
  year = {2022},
  month = {Dec},
  publisher = {American Physical Society},
  doi = {10.1103/PhysRevLett.129.267701},
  url = {https://link.aps.org/doi/10.1103/PhysRevLett.129.267701}
}

@article{qiao2025size,
  title={Size optimization for observeing Majorana fermions},
  author={Qiao, Guo-Jian and Zhang, Zhi-Lei and Yue, Xin and Sun, CP},
  journal={arXiv preprint arXiv:2511.21423},
  year={2025}
}

@article{yue2025finite,
  title = {Finite-size effects on metallization versus chiral Majorana fectmions},
  author = {Yue, Xin and Qiao, Guo-Jian and Sun, C. P.},
  journal = {Phys. Rev. B},
  volume = {113},
  issue = {11},
  pages = {115416},
  numpages = {9},
  year = {2026},
  month = {Mar},
  publisher = {American Physical Society},
  doi = {10.1103/57w4-rs6c},
  url = {https://link.aps.org/doi/10.1103/57w4-rs6c}
}

@article{van2026single,
  title={Single-shot parity readout of a minimal Kitaev chain},
  author={van Loo, Nick and Zatelli, Francesco and Steffensen, Gorm O and Roovers, Bart and Wang, Guanzhong and Van Caekenberghe, Thomas and Bordin, Alberto and van Driel, David and Zhang, Yining and Huisman, Wietze D and others},
  journal={Nat.},
  volume={650},
  number={8101},
  pages={334--339},
  year={2026},
  publisher={Nature Publishing Group UK London},
  url = {https://www.nature.com/articles/s41586-025-09927-7}
}

@article{Wang_2015,
  title = {Chiral topological superconductor and half-integer conductance plateau from quantum anomalous Hall plateau transition},
  author = {Wang, Jing and Zhou, Quan and Lian, Biao and Zhang, Shou-Cheng},
  journal = {Phys. Rev. B},
  volume = {92},
  issue = {6},
  pages = {064520},
  numpages = {8},
  year = {2015},
  month = {Aug},
  publisher = {American Physical Society},
  doi = {10.1103/PhysRevB.92.064520},
  url = {https://link.aps.org/doi/10.1103/PhysRevB.92.064520}
}

@article{Chung_2011,
  title = {Conductance and noise signatures of Majorana backscattering},
  author = {Chung, Suk Bum and Qi, Xiao-Liang and Maciejko, Joseph and Zhang, Shou-Cheng},
  journal = {Phys. Rev. B},
  volume = {83},
  issue = {10},
  pages = {100512(R)},
  numpages = {4},
  year = {2011},
  month = {Mar},
  publisher = {American Physical Society},
  doi = {10.1103/PhysRevB.83.100512},
  url = {https://link.aps.org/doi/10.1103/PhysRevB.83.100512}
}

@article{2025PRB,
  title = {Non-Majorana origin of the half-integer conductance quantization elucidated by multiterminal superconductor--quantum anomalous Hall insulator heterostructure},
  author = {Uday, Anjana and Lippertz, Gertjan and Bhujel, Bibek and Taskin, Alexey A. and Ando, Yoichi},
  journal = {Phys. Rev. B},
  volume = {111},
  issue = {3},
  pages = {035440},
  numpages = {10},
  year = {2025},
  month = {Jan},
  publisher = {American Physical Society},
  doi = {10.1103/PhysRevB.111.035440},
  url = {https://link.aps.org/doi/10.1103/PhysRevB.111.035440}
}

@article{bordin2025enhanced,
  title={Enhanced Majorana stability in a three-site Kitaev chain},
  author={Bordin, Alberto and Liu, Chun-Xiao and Dvir, Tom and Zatelli, Francesco and Ten Haaf, Sebastiaan LD and Van Driel, David and Wang, Guanzhong and Van Loo, Nick and Zhang, Yining and Wolff, Jan Cornelis and others},
  journal={Nat. Nanotechnol.},
  volume={20},
  number={6},
  pages={726--731},
  year={2025},
  publisher={Nature Publishing Group UK London},
  url = {https://www.nature.com/articles/s41565-025-01894-4#citeas}
}

@article{bordin2026probing,
  title={Probing Majorana localization of a phase-controlled three-site Kitaev chain with an additional quantum dot},
  author={Bordin, Alberto and Bennebroek Evertsz’, Florian J and Roovers, Bart and Torres Luna, Juan D and Huisman, Wietze D and Zatelli, Francesco and Mazur, Grzegorz P and Ten Haaf, Sebastiaan LD and Badawy, Ghada and Bakkers, Erik PAM and others},
  journal={Nat. Commun.},
  volume={17},
  pages={2313},
  year={2026},
  publisher={Nature Publishing Group UK London},
  url = {https://www.nature.com/articles/s41467-026-68897-0?fromPaywallRec=false#citeas}
}

@article{ten2025observation,
  title={Observation of edge and bulk states in a three-site Kitaev chain},
  author={Ten Haaf, Sebastiaan LD and Zhang, Yining and Wang, Qingzhen and Bordin, Alberto and Liu, Chun-Xiao and Kulesh, Ivan and Sietses, Vincent PM and Prosko, Christian G and Xiao, Di and Thomas, Candice and others},
  journal={Nature},
  volume={641},
  number={8064},
  pages={890--895},
  year={2025},
  publisher={Nature Publishing Group UK London},
  url = {https://www.nature.com/articles/s41586-025-08892-5}
}

@article{PhysRevB.81.241310,
  title = {Proximity effect at the superconductor--topological insulator interface},
  author = {Stanescu, Tudor D. and Sau, Jay D. and Lutchyn, Roman M. and Das Sarma, S.},
  journal = {Phys. Rev. B},
  volume = {81},
  issue = {24},
  pages = {241310},
  numpages = {4},
  year = {2010},
  month = {Jun},
  publisher = {American Physical Society},
  doi = {10.1103/PhysRevB.81.241310},
  url = {https://link.aps.org/doi/10.1103/PhysRevB.81.241310}
}

@article{PhysRevB.96.014510,
  title = {Proximity-induced low-energy renormalization in hybrid semiconductor-superconductor Majorana structures},
  author = {Stanescu, Tudor D. and Das Sarma, Sankar},
  journal = {Phys. Rev. B},
  volume = {96},
  issue = {1},
  pages = {014510},
  numpages = {19},
  year = {2017},
  month = {Jul},
  publisher = {American Physical Society},
  doi = {10.1103/PhysRevB.96.014510},
  url = {https://link.aps.org/doi/10.1103/PhysRevB.96.014510}
}

@article{PhysRevB.84.144522,
  title = {Majorana fermions in semiconductor nanowires},
  author = {Stanescu, Tudor D. and Lutchyn, Roman M. and Das Sarma, S.},
  journal = {Phys. Rev. B},
  volume = {84},
  issue = {14},
  pages = {144522},
  numpages = {29},
  year = {2011},
  month = {Oct},
  publisher = {American Physical Society},
  doi = {10.1103/PhysRevB.84.144522},
  url = {https://link.aps.org/doi/10.1103/PhysRevB.84.144522}
}

@article{PhysRevLett.106.127001,
  title = {Search for Majorana Fermions in Multiband Semiconducting Nanowires},
  author = {Lutchyn, Roman M. and Stanescu, Tudor D. and Das Sarma, S.},
  journal = {Phys. Rev. Lett.},
  volume = {106},
  issue = {12},
  pages = {127001},
  numpages = {4},
  year = {2011},
  month = {Mar},
  publisher = {American Physical Society},
  doi = {10.1103/PhysRevLett.106.127001},
  url = {https://link.aps.org/doi/10.1103/PhysRevLett.106.127001}
}

@article{alvarado2026optimal,
  title={Optimal Majoranas in Mesoscopic Kitaev Chains},
  author={Alvarado, M and Souto, R Seoane and Calder{\'o}n, Mar{\'\i}a Jos{\'e} and Aguado, Ram{\'o}n},
  journal={arXiv preprint arXiv:2604.13945},
  year={2026}
}
\end{document}